\documentclass[11pt]{article}
\pdfoutput=1
\usepackage{amsmath,amssymb,amsfonts}
\usepackage[nosort]{cite}
\usepackage{graphicx}
\usepackage[colorlinks=true, pdfstartview=FitV, linkcolor=blue, citecolor=magenta, urlcolor=purple]{hyperref}
\usepackage{mdframed,framed}
\usepackage{cancel}
\usepackage[usenames,dvipsnames]{xcolor}
\usepackage{notoccite}
\newcommand{\beq}{\begin{equation}}
\newcommand{\eeq}{\end{equation}}
\newcommand{\beqn}{\begin{eqnarray}}
\newcommand{\eeqn}{\end{eqnarray}}
\newcommand{\pa}{\partial}

\usepackage{cancel}

\newcommand{\un}[1]{\underline{#1}}

\newcommand{\comment}[1]{}

\newcommand{\D}{\text{d}}

\usepackage{databib}

\usepackage{tocloft}
\newlength{\leftfield}          
\newlength{\centerfield}        
\newlength{\rightfield}         
\newcommand{\twodigits}[1]{\ifnum#1<10 0\fi\number#1}
	
\newcommand{\mytitlebox}[4]{
	\framebox[\textwidth]{
		\begin{minipage}[c]{\textwidth}
		\makebox[\leftfield][l]{\littlefig{Physics-logo-small.png}{3} 
		}\ 
		\begin{minipage}[b]{\centerfield}\begin{center}{\bf\large #1}\\ #2\end{center}\end{minipage} 
		\begin{minipage}[b]{\rightfield}\begin{flushright}#3\\ Create: #4\\ Update: \the\year--\twodigits\month--\twodigits\day \end{flushright}\end{minipage} 
		\end{minipage}
	}
}	

\author{
  \begin{minipage}{.97\linewidth}
    \vspace{1cm}
       \begin{center}
      \begin{small}  
               \textbf{Francesco Alessio},$^{a,b}$ 
             \textbf{Glenn Barnich},$^b$ 
     \textbf{Luca Ciambelli}$^b$, 
     \textbf{Pujian Mao}$^c$ and
      \textbf{Romain Ruzziconi}$^{d}$     
              \end{small}
    \end{center}
    \vspace{0.5cm}
      \hspace{2.4cm}\begin{minipage}{.7\linewidth}
\begin{center}     {\it \begin{footnotesize}
\hbox{
\kern-2cm
\vbox{
\begin{itemize}
 \item[$^a$]
Dipartimento di Fisica ``E. Pancini" and INFN,\\ Universit\`a degli studi di Napoli ``Federico II",\\ I-80125 Napoli, Italy 
      \end{itemize}
      \vskip0.20cm
      }   
\kern-2.8cm
      \vbox{
 \begin{itemize}
                  \item[$^b$]
Universit\'e Libre de Bruxelles \\and International Solvay Institutes\\
CP 231, B-1050 Brussels, Belgium
      \end{itemize}}
    }
    \hbox{
\kern-2cm
\vbox{
\begin{itemize}
 \item[$^c$]
Center for Joint Quantum Studies and Department of Physics,\\ School of Science, Tianjin University,\\135 Yaguan Road, Tianjin 300350, China
      \end{itemize}
      \vskip0.20cm
      }   
\kern-2.8cm
      \vbox{
 \begin{itemize}
                  \item[$^d$]
Institute for Theoretical Physics,\\ TU Wien,\\Wiedner Hauptstrasse 8, A-1040 Vienna, Austria
      \end{itemize}
        \vskip0.20cm}
    }
     \end{footnotesize}}
\end{center}
    \end{minipage}
      \vspace{0.5cm}
  \end{minipage}
}
\title{\vspace{1.5cm}
 \boldmath 
    \textbf{Weyl Charges in Asymptotically Locally AdS$_3$ Spacetimes}
  \unboldmath
}

\date{}

\begin{document}

\begin{titlepage}
\maketitle
\thispagestyle{empty}

\begin{center}
\textsc{Abstract}\\  
\vspace{1cm}	
\begin{minipage}{1.0\linewidth}

We discuss an enhancement of the Brown--Henneaux boundary conditions in three-dimensional AdS General Relativity to encompass Weyl transformations of the boundary metric. The resulting asymptotic symmetry algebra, after a field-dependent redefinition of the generators, is a direct sum of two copies of the Witt algebra and the Weyl abelian sector. The charges associated to Weyl transformations are non-vanishing, integrable but not conserved due to a flux driven by the Weyl anomaly coefficient. The charge algebra admits an additional non-trivial central extension in the Weyl sector, related to the well-known Weyl anomaly. We then construct the holographic Weyl current and show that it satisfies an anomalous Ward--Takahashi identity of the boundary theory.

\end{minipage}
\end{center}

\end{titlepage}

\begingroup
\hypersetup{linkcolor=black}
\tableofcontents
\endgroup
\noindent\rule{\textwidth}{0.6pt}
\newpage

\section{Introduction}

Three-dimensional General Relativity is one of the simplest gravitational systems \cite{Deser:1983tn,Deser:1983nh} and, in particular, solutions with negative cosmological constant (AdS${}_3$) have received special attention, due to their holographic nature \cite{Brown:1986nw, Witten:1998qj}. The absence of bulk propagating degrees of freedom makes this theory a privileged playground to better understand the role of boundary conditions in gravity. Indeed, the dynamics can be  described by a pure boundary theory, as shown in the Chern--Simons formulation \cite{Achucarro:1987vz, Witten:1988hc,Banados:1994tn,Coussaert:1995zp,Henneaux:1999ib,Allemandi:2002sx,Banados:2006fe,Banados:2016zim}, (for a recent review, see also \cite{Donnay:2016iyk}).

Boundary conditions play a pivotal role in physics. Together with the choice of a bulk gauge for the metric, they fully determine the field content -- the solution space -- of the theory. Residual diffeomorphisms are those preserving the gauge choice. Among them, the ones respecting boundary conditions and carrying non-vanishing surface charges are the so-called asymptotic symmetry generators \cite{Regge:1974zd,Benguria:1976in,Brown:1986nw,Barnich:2001jy,Carlip:2005tz,Barnich:2007bf}.\footnote{For recent reviews, see \cite{Oblak:2016eij,Compere:2018aar,Ruzziconi:2019pzd}.}  The surface charges are interesting quantities, for they encode observables of the system, such as its energy and momenta \cite{Wald:1993nt,Iyer:1994ys}.\footnote{Recently there has been a renewed interest in the charges structure of spacetime corners
\cite{Freidel:2020xyx,Freidel:2020svx,Freidel:2020ayo}.} The asymptotic symmetry generators form the asymptotic symmetry algebra, represented on the solution space by the projective charge algebra, trustworthy up to a  universal central extension \cite{Brown:1986nw}. Probing various boundary conditions and their related surface charges is a natural question, the literature on the topic is extensive, see e.g. \cite{Barnich:2006av,Compere:2008us,Compere:2013bya,Troessaert:2013fma,Avery:2013dja,Grumiller:2016pqb,Grumiller:2017sjh,Poole:2018koa,Henneaux:2019sjx,Bergshoeff:2019rdb,Donnay:2020fof,Adami:2020ugu}. Along this line of thought, in this work we introduce a new set of boundary conditions, justified below, and study its consequences.

In the seminal work by Brown and Henneaux (BH) \cite{Brown:1986nw} it was shown that the asymptotic symmetry algebra of AdS$_3$, under certain boundary conditions encompassing BTZ black holes \cite{Banados:1992wn,Banados:1992gq,Carlip:1995qv}, consists in two commuting copies of the Virasoro algebra with central extensions $c^{\pm}=\frac{3\ell}{2G}$, $\ell$ being the AdS${}_3$ radius and $G$ the Newton constant. This result is considered as a precursor of the AdS/CFT correspondence \cite{tHooft:1993dmi, Susskind:1994vu, Maldacena:1997re, Gubser:1998bc, Witten:1998qj}, which, applied to three-dimensional General Relativity, conjectures the existence of a dual Confomal Field Theory (CFT) living on the two-dimensional boundary.  Remarkably, the value of $c^{\pm}$ has been used to microscopically derive the Bekenstein--Hawking  entropy of the BTZ black hole \cite{Strominger:1997eq}, using the Cardy formula \cite{Cardy:1986ie}. Moreover, by taking a suitable flat limit of asymptotically AdS${}_3$ gravity \cite{Barnich:2012aw}, it is possible to extend these considerations to asymptotically flat spacetimes \cite{Barnich:2010eb,Campoleoni:2018ltl}.   

In the context of Penrose conformal compactification \cite{Penrose:1962ij,Penrose1964}, applied to the case of  AdS$_3$ spacetime, the bulk metric induces a boundary conformal class $\left[g^{(0)}\right]$ of metrics rather than a metric \cite{FG1,Henningson:1998gx,Skenderis:1999nb,Rooman:2000zi,Rooman:2000ei,Fefferman:2007rka,Alessio:2017lps,Ciambelli:2019bzz}. The boundary conditions considered here are motivated by this approach. In BH, a particular representative of the equivalence class is  picked up, namely the flat Minkowski metric $\eta$, and kept fixed under the action of the asymptotic symmetry algebra. This  defines asymptotically (globally) AdS${}_3$ spacetimes (AAdS${}_3$). 

In this manuscript we focus on asymptotically locally AdS${}_3$ (AlAdS${}_3$) spacetimes \cite{Graham:1991jqw,Graham:1999jg,Papadimitriou:2005ii,Fischetti:2012rd},  with no restriction on their boundary conformal structure. In general, the two-dimensional boundary metric $g^{(0)}$ is specified by three arbitrary functions in terms of which the Einstein equations can be exactly solved. We work in the Fefferman--Graham (FG) gauge \cite{FG1,Fefferman:2007rka} and we assume the boundary metric to be conformally flat, the conformal factor being an arbitrary smooth function independent of the radial coordinate.\footnote{The case in which the conformal factor admits a chiral splitting has been extensively analysed in previous works \cite{Troessaert:2013fma,Barnich2014TheDT}.} The resulting asymptotic symmetries contain the usual two copies of diffeomorphisms of the circle together with additional  Weyl transformations of the boundary metric. These are often referred to in the literature as Penrose--Brown--Henneaux (PBH)\cite{Imbimbo:1999bj} transformations. Here we explicitly compute their associated surface charges,\footnote{For the surface charges we use the prescription given in \cite{Barnich:2001jy}.} and find that they are finite, integrable but non conserved, which is an interesting unusual combination (see \cite{Barnich:2007bf,Adami:2020ugu,Chandrasekaran:2020wwn} for related discussions).

Diffeomorphisms generating boundary Weyl rescalings are crucial in the context of holographic renormalization, pioneered by Skenderis and collaborators \cite{Henningson:1998gx,deHaro:2000vlm,Bianchi:2001kw,Skenderis:2002wp,Papadimitriou:2004ap,Papadimitriou:2005ii} (see also \cite{Emparan:1999pm,Kalkkinen:2001vg,Banados:2005rz,Compere:2008us,Anastasiou:2020zwc}). Regularizing the theory explicitly breaks Weyl invariance causing the emergence of a Weyl anomaly \cite{Henningson:1998gx,Imbimbo:1999bj,Bautier:1999ic,Schwimmer:2000cu,Ciambelli:2019bzz}.\footnote{For intrinsic field-theoretical studies of Weyl anomalies see \cite{Fradkin:1983tg,Duff:1993wm,Deser:1993yx,Deser:1996na,Boulanger:2007ab,Schwimmer:2008yh,Adam:2009gq,Schwimmer:2010za}.} The latter can be seen in the on-shell variational principle of the renormalized bulk action.
When specified to a variation of the conformal factor of the boundary metric, the corresponding variation of the on-shell action gives the Weyl anomaly, which is then interpreted as the trace anomaly of the boundary stress tensor \cite{DiFrancesco:1997nk,Balasubramanian:1999re,Skenderis:2000in}. Typically, in order to achieve a well-defined variational problem, Dirichlet boundary conditions are imposed on the metric \cite{Regge:1974zd,Papadimitriou:2005ii,Compere:2008us}. However, such a condition is too restrictive when working with a conformal class of boundary metrics \cite{Papadimitriou:2005ii}. Therefore we cannot insist that the variational problem be well defined and we impose only the cocycle condition on the second Weyl variation of the on-shell action \cite{Bonora:1985cq,Karakhanian:1994yd,Mazur:2001aa,Manvelyan:2001pv,Boulanger:2007st}. It turns out that the Weyl surface charges are finite and integrable, whereas their non conservation is accounted for by a symplectic flux through the boundary \cite{Wald:1993nt,Iyer:1994ys,Wald:1999wa,Harlow:2019yfa,Compere:2019bua,Compere:2020lrt}. The presence of an anomaly indicates that, in the dual theory, a current is not conserved at the quantum level, see \cite{Cheng:1985bj,Treiman:1986ep,Bilal:2008qx} for reviews. We construct new Weyl boundary currents compatible with the surface charges. Their non conservation translates into the anomalous Ward--Takahashi identity \cite{Ward:1950xp,Takahashi:1957xn} associated to Weyl symmetry of the putative holographic theory.

The paper is organized as follows. In Section \ref{S0}, we fix the FG gauge and introduce conformally flat boundary conditions.  Correspondingly, we compute the asymptotic Killing vectors preserving these choices and their algebra. We show that the latter comprises, besides the usual left and right Witt sectors, a new abelian sector corresponding to Weyl rescalings of the boundary metric. We then solve Einstein equations and extract the action of the asymptotic symmetry algebra on solution space. In Section \ref{S2}, we compute the corresponding surface charges. Furthermore, we show that the charge algebra is centrally extended in both the Witt and the Weyl sector. In Section \ref{sec4}, we touch upon some features of the boundary holographic theory. We show in detail that, under our choice of boundary conditions, the variational problem is not well-defined due to the presence of the Weyl anomaly. We construct the boundary Weyl currents and show that their non-conservation can be interpreted in terms of an anomalous Ward-Takahashi identity for the boundary Weyl transformations. We close in Section \ref{S6} with a short summary and perspectives. Appendix \ref{AppA} contains a brief comparison of this work with \cite{Troessaert:2013fma}, whilst in Appendix \ref{AppB} we translate our results to the Chern--Simons formulation of the theory. 

\section{New Boundary Conditions}
\label{S0}

The new boundary conditions considered in this work are motivated by the geometric approach originally introduced by Penrose \cite{Penrose:1962ij,Penrose1964} in the context of conformal compactification. In this framework,  the boundary data for the full metric $g$ are located at infinite distance, due to the second order pole structure typical of AdS. Multiplying $g$ by $\Omega^2$, with $\Omega$ a positive function with a simple zero on the boundary, such a pole is eliminated and an induced metric on the boundary may be defined. There is however an ambiguity in the choice of $\Omega$. The replacement $\Omega\rightarrow \Omega'=e^{\omega}\Omega$, with $\omega$ a smooth function independent of the radial coordinate, induces a conformal transformation  $g^{(0)}\rightarrow g^{(0)}_{{}_{\omega}}= e^{2\omega}g^{(0)}$ of the boundary metric. Such a freedom allows one to define only an equivalence class of conformally related boundary metrics, $\left[g^{(0)}\right]$, rather than a metric \cite{Rooman:2000zi,Rooman:2000ei,FG1,Fefferman:2007rka,Alessio:2017lps,Ciambelli:2019bzz}.

\subsection{Fefferman--Graham and Residual Diffeomorphisms}
\label{S1}
The FG gauge \cite{FG1, Fefferman:2007rka}\footnote{For a recent discussion see also \cite{Ruzziconi:2019pzd, Ciambelli:2020ftk, Ciambelli:2020eba}.} in three spacetime dimensions consists in choosing coordinates $x^{\mu}=(\rho,x^a)$, where $\rho\geq 0$ is a radial coordinate and $x^a=(t,\phi)$. The three gauge-fixing conditions for the metric are $g_{\rho\rho}=\frac{\ell^2}{\rho^2}$ and $g_{\rho a}=0$, where $\ell^2=-\frac{1}{\Lambda}$ is the AdS${}_3$ radius. The boundary is located at $\rho=0$. The line element takes the form
\begin{align}
\label{FG}
\D s^2=g_{\mu\nu}\text dx^\mu \text dx^\nu=\frac{\ell^2}{\rho^2}\D\rho^2+\gamma_{ab}(\rho,x)\D x^a\D x^b.
\end{align}
We solve Einstein equations with initial boundary condition $
\gamma_{ab}(\rho,x)=\mathcal{O}(\rho^{-2})$, so that $\Omega^2 g_{\mu\nu}$ is well defined at the boundary.  Einstein equations for \eqref{FG} yield
\begin{align}
\label{FGExp}
\gamma_{ab}(\rho,x)=\rho^{-2}g^{(0)}_{ab}(x)+g^{(2)}_{ab}(x)+\rho^2g^{(4)}_{ab}(x),
\end{align}
with
\begin{align}
\label{Terms}
g^{(4)}_{ab}=\frac{1}{4}g_{ac}^{(2)}g^{cd}_{(0)}g_{db}^{(2)},\hspace{1cm}g_{(0)}^{ab}g^{(2)}_{ab}=-\frac{\ell^2}{2}R^{(0)},\hspace{1cm}D^a_{(0)}g_{ab}^{(2)}=-\frac{\ell^2}{2}\partial_bR^{(0)}.
\end{align}
We denote by $R^{(0)}$ and $D^{(0)}_a$ the Ricci scalar and the covariant derivative associated to $g_{ab}^{(0)}$, respectively. The leading term  $g_{ab}^{(0)}$ of the expansion \eqref{FGExp} as $\rho\rightarrow 0$ is usually referred to as the boundary metric. From now on the indices will be raised and lowered using this metric.

Defining the holographic stress-energy tensor \cite{Henningson:1998gx, Balasubramanian:1999re} as
\begin{align}
\label{T}
T_{ab}=\frac{1}{8\pi G\ell}\left(g^{(2)}_{ab}+\frac{\ell^2}{2}g^{(0)}_{ab}R^{(0)}\right), 
\end{align}
the last two equations of \eqref{Terms} imply
\begin{align}
\label{ST}
T_{a}{}^{a}=\frac{c}{24\pi}R^{(0)},\hspace{1cm}D_a^{(0)}T^{ab}=0,
\end{align}
where $c=\frac{3\ell}{2G}$ is the BH central charge \cite{Brown:1986nw}. The first equation in \eqref{ST} states that, for a general $g^{(0)}_{ab}$ whose Ricci scalar is non-vanishing, the trace of the tensor $T_{ab}$ defined in \eqref{T} is non-vanishing and proportional to the scalar curvature $R^{(0)}$, with a proportionality constant that is determined by the BH central charge. Hence the dual CFT living on the boundary must have a Weyl anomaly. We will further comment on this in Section \ref{sec4}. The full solution space $\chi$ is therefore characterized by five functions, three contained in $g^{(0)}_{ab}$ and two in $g^{(2)}_{ab}$ or, equivalently, in $T_{ab}$. Furthermore, these last two functions satisfy the dynamical constraints \eqref{Terms} or, equivalently, the second equation in \eqref{ST}. In the following, we will write $\chi=\{g^{(0)}_{ab},g^{(2)}_{ab}\}$.

The residual gauge diffeomorphisms are those preserving the FG gauge conditions. They are thus generated by the vector $\un\xi$ satisfying
\begin{align}
\label{2}
\mathcal{L}_{\un \xi}g_{\rho\rho}=0,\qquad \mathcal{L}_{\un \xi}g_{\rho a}=0, \qquad \mathcal{L}_{\un \xi}\gamma_{ab}=\mathcal{O}(\rho^{-2}).
\end{align}
The solution of these equations is
\beq
\un\xi:= \xi^\mu\pa_\mu=\xi^\rho\pa_\rho+\xi^a\pa_a,\label{AKV}
\eeq
with
\begin{align}
\label{3}
\xi^{\rho}=\rho\sigma(x),\hspace{1cm}\xi^a=Y^a(x)-\ell^2\partial_b\sigma(x)\int_{0}^{\rho}\frac{\D \rho'}{\rho'}\gamma^{ab}(\rho',x).
\end{align}
In this expression, $\sigma(x)$ and $Y^a(x)$ are field-independent arbitrary functions and we note that $\xi^a$ depends on the metric field $\gamma^{ab}$. This motivates the introduction of the modified Lie bracket \cite{Barnich:2010eb}
\begin{align}
\label{4}
\big[\un \xi_1,\un\xi_2\big]_{M}:=\big[\un\xi_1,\un\xi_2\big]-\delta_{\un\xi_1}\un\xi_2+\delta_{\un\xi_2}\un\xi_1,
\end{align}
to study the asymptotic algebra. Here $\big[\cdot,\cdot\big]$ denotes the ordinary Lie bracket between vector fields and $\delta_{\un\xi_1}\un\xi_2[g]$ the variation of the vector field $\un\xi_2[g]$ due to the metric variation $\delta_{\un\xi}g=\mathcal{L}_{\un\xi}g$, i.e. $\delta_{\un\xi_1}\un\xi_2[g]=\un\xi_2[\delta_{\un\xi_1}g]$. 
On defining
\beq
\label{6}
\hat{\xi}^{\rho}=\rho\hat{\sigma}(x), \qquad \hat{\sigma}(x)=Y_1^a(x)\partial_a\sigma_2(x)-Y_2^a(x)\partial_a\sigma_1(x),
\eeq
and
\beq
\hat{\xi}^{a}=\hat{Y}^a(x)-\ell^2\partial_{b}\hat\sigma(x)\int_{0}^{\rho}\frac{\D\rho'}{\rho'}{\gamma}^{ab}(\rho',x),\qquad\hat{Y}^a(x)=Y_1^b(x)\partial_bY_2^a(x)-Y_2^b(x)\partial_bY_1^a(x),
\eeq
it is possible to show that our algebra is closed off shell \cite{Barnich:2010eb,Compere:2020lrt}
\begin{align}
\label{5}
\big[\un\xi_1,\un\xi_2\big]_{M}=\hat{\un\xi}.
\end{align}
To prove this we used that $\big[\xi_1,\xi_2\big]^a_{M}$, i.e. the $a$ component of $\big[\un\xi_1,\un\xi_2\big]_{M}=\big[\xi_1,\xi_2\big]^\rho_{M}\pa_\rho+\big[\xi_1,\xi_2\big]^a_{M}\pa_a$, satisfies the differential equation $\partial_{\rho}\big[\xi_1,\xi_2\big]^a_{M}=-\frac{\ell^2}{\rho^2}\gamma^{ab}\partial_b\big[\xi_1,\xi_2\big]^{\rho}_{M}$ with boundary condition $\lim_{\rho\to 0}\big[\xi_1,\xi_2\big]^a_{M}=\hat{Y}^a$.
On shell, the residual diffeomorphism generator \eqref{3} becomes 
\begin{align}
\label{12}
\xi^a=Y^a-\frac{\rho^2}{2}\ell^2g^{ab}_{(0)}\partial_{b}\sigma+\frac{\rho^4}{4}\ell^2g^{ac}_{(0)}g^{(2)}_{cd}g^{db}_{(0)}\partial_b\sigma+\mathcal{O}(\rho^{6}).
\end{align}

Acting with the Lie derivative along $\un\xi$ on the on-shell line element \eqref{FG} we find the general variation of solution space
\begin{align}
\label{13}
\left(\mathcal{L}_{\un \xi}g_{\mu\nu}\right)\D x^{\mu}\D x^{\nu}=\frac{\ell^2}{\rho^2}\D\rho^2+\left(\rho^{-2}\delta_{\un \xi}g^{(0)}_{ab}+\delta_{\un \xi}g^{(2)}_{ab}+\rho^2\delta_{\un \xi}g^{(4)}_{ab}\right)\D x^a\D x^b,
\end{align}
with
\beqn
\label{14}
&\delta_{\un \xi}g^{(0)}_{ab}=\mathcal{L}_{\un Y}g^{(0)}_{ab}-2\sigma g^{(0)}_{ab},\qquad \delta_{\un \xi}g^{(2)}_{ab}=\mathcal{L}_{\un Y}g^{(2)}_{ab}-\ell^2D^{(0)}_aD^{(0)}_b\sigma.
\eeqn
The first equation in \eqref{14} is telling us that a general variation of the boundary metric under the action of residual gauge diffeomorphisms has two independent contributions, one coming from $\sigma$ and the other from $Y^a$.

\subsection{Boundary Gauge Fixing}\label{SS4}

As stressed above, once a boundary metric $g^{(0)}_{ab}$ is assigned, the full solution space, comprising also the two functions in $g^{(2)}_{ab}$, is completely determined. That is, for every arbitrary choice of the boundary metric, solving \eqref{Terms} yields a complete solution of Einstein equations. However, as already advertised in the introduction, we do not leave $g^{(0)}_{ab}$ arbitrary, but impose the boundary condition
\beq
g_{ab}^{(0)}(x)=e^{2\varphi(x)}\eta_{ab},\label{confg}
\eeq
where $\eta_{ab}$ is the $2$-dimensional Minkowski metric in Lorentzian signature. Notice that every $2$-dimensional metric is conformally flat. That is, we can use boundary diffeomorphisms to fix $2$ components of the boundary metric in order to reach \eqref{confg}. This condition will constrain the form of the vector fields $Y^a$ appearing in \eqref{3}. Although \eqref{confg} is a restrictive boundary condition, it is a natural case to investigate. Note that an arbitrary variation of the boundary metric now reduces to an arbitrary variation of its conformal factor, i.e. $\delta g^{(0)}_{ab}=2(\delta\varphi )g^{(0)}_{ab}$.

Eq. \eqref{14} becomes then
\begin{align}
\label{16}
\delta_{\un \xi}g^{(0)}_{ab}=\mathcal{L}_{\un Y}g_{ab}^{(0)}-2\sigma g^{(0)}_{ab}=2(\delta_{\un\xi}\varphi)g_{ab}^{(0)}.
\end{align}
This means that $\un Y$ is a conformal Killing vector of $g^{(0)}_{ab}$
\begin{align}
\label{17}
\mathcal{L}_{\un Y}g^{(0)}_{ab}=D^{(0)}_{a}Y_b+D^{(0)}_{b}Y_a=2\Omega_{\un Y} g^{(0)}_{ab}, \qquad \Omega_{\un Y}=\frac{1}{2}D^{(0)}_aY^a.
\end{align}
where $\Omega_{\un Y}=\delta_{\un\xi}\varphi+\sigma$. Thence
\begin{align}
\label{16}
\delta_{\un \xi}g^{(0)}_{ab}=2(\Omega_{\un Y}-\sigma) g^{(0)}_{ab}.
\end{align}
Introducing light-cone coordinates $x^{\pm}=\frac{t}{\ell}\pm\phi$ we have $g^{(0)}_{ab}\D x^a\D x^b=-e^{2\varphi(x^+,x^-)}\D x^+\D x^-$ and \eqref{17} is solved by the usual chiral vectors
\begin{align}
\label{18}
Y^{+}=Y^+(x^+),\qquad Y^-=Y^-(x^-),\qquad \Omega_{\un Y}=\frac{1}{2}\left(\partial_-Y^-+\partial_+Y^+\right)+Y^+\partial_+\varphi+Y^-\partial_-\varphi.
\end{align}
Consistently, the only effect of the residual gauge symmetries on the boundary metric is to induce a Weyl transformation, i.e. a shift in its conformal factor, given by $\delta_{\un \xi}\varphi=\Omega_{\un Y}-\sigma$.

The standard Brown--Henneaux boundary conditions \cite{Brown:1986nw} $\delta_{\xi}\varphi=0$  are a subclass of our boundary conditions obtained by imposing $\sigma=\Omega_{\un Y}$. With this choice the effect of the conformal isometry generated by $\un Y$ exactly compensates the effect of the Weyl rescaling due to $\sigma$, as clear from \eqref{14}. Furthermore, also the boundary conditions studied in  \cite{Troessaert:2013fma} are encompassed in our analysis, as we show in Appendix \ref{AppA}.

\subsection{Solution Space}\label{SS5}

In the conformally flat parametrization it is possible to  explicitly solve Einstein equations for $g^{(2)}_{ab}$ given by the last two equations of \eqref{Terms}, \cite{Barnich:2010eb}. The first is an algebraic equation for $g^{(2)}_{+-}$ and yields
\begin{align}
\label{19}
g^{(2)}_{+-}=\ell^2\partial_{+}\partial_-\varphi,
\end{align}
where we used $R^{(0)}=8e^{-2\varphi}\partial_+\partial_-\varphi$. The second implies
\begin{align}
\label{20}
&\partial_{\mp}g^{(2)}_{\pm\pm}=-\ell^2\left(2\partial_{\pm}\varphi\partial_{\pm}\partial_{\mp}\varphi-\partial^2_{\pm}\partial_{\mp}\varphi\right),
\end{align}
with solutions
\begin{align}
\label{21}
g^{(2)}_{\pm\pm}=\ell^2\left[\Xi_{\pm\pm}(x^{\pm})+\partial^2_{\pm}\varphi-(\partial_{\pm}\varphi)^2\right],
\end{align}
where $\Xi_{\pm\pm}(x^{\pm})$ are two arbitrary functions of $x^{\pm}$. The holographic energy-momentum tensor \eqref{T} is 
\begin{align}
\label{22}
T_{+-}=-\frac{\ell}{8\pi G}\partial_+\partial_-\varphi,\qquad T_{\pm\pm}=\frac{\ell}{8\pi G}\left[\Xi_{\pm\pm}(x^{\pm})+\partial_{\pm}^2\varphi-(\partial_{\pm}\varphi)^2\right].
\end{align}
While the general solution space is characterized by five independent functions of $x^+$ and $x^-$, the solution space in the conformally flat gauge is given by $\varphi(x^+,x^-)$ and the two chiral functions $\Xi_{\pm\pm}(x^{\pm})$. Thus, the solution space is  $\chi=\{\Xi_{++}(x^+),\Xi_{--}(x^-),\varphi(x^+,x^-)\}$. Note that the presence of an arbitrary $\varphi(x^+,x^-)$ prevents a complete chiral splitting of the solution space and that, equivalently, the holographic stress-energy tensor components $T_{\pm\pm}$ in \eqref{22} are not chiral nor antichiral. This is one of the main differences with respect to \cite{Troessaert:2013fma}.

A generic variation of the solution space is generated by $\sigma$ and $Y^\pm$, so we symbolically write $\delta_{\un \xi}\chi=\delta_{(\sigma,Y^\pm)}\chi$. Using \eqref{14} with \eqref{17} we compute
\begin{align}
\label{23}
\delta_{(\sigma,0)}\varphi=-\sigma,\qquad\delta_{(\sigma,0)}\Xi_{\pm\pm}=0,
\end{align}
and
\begin{align}
\label{24}
\delta_{(0,Y^\pm)}\varphi=\partial_-Y^-+\partial_+Y^++2(Y^+\partial_+\varphi+Y^-\partial_-\varphi),\quad \delta_{(0,Y^\pm)}\Xi_{\pm\pm}=Y^{\pm}\partial_{\pm}\Xi_{\pm\pm}+2\Xi_{\pm\pm}\partial_{\pm}Y^{\pm}-\frac{1}{2}\partial^3_{\pm}Y^{\pm}.
\end{align}
Before proceeding to calculate the asymptotic symmetry algebra, it is convenient to trade $\sigma(x^+,x^-)$ for the new field dependent parameter $\omega(x^+,x^-)$ as
\begin{align}
\label{24.1}
\omega=\Omega_{\un Y}-\sigma.
\end{align}
Note that $\omega$ depends on the derivatives of $\varphi$. Using $\omega$, eqs. (\ref{23}-\ref{24}) can be more compactly written as
\begin{align}
\label{24.2}
\delta_{(\omega, Y^\pm)}\varphi=\omega,\qquad \delta_{(\omega, Y^\pm)}\Xi_{\pm\pm}=Y^{\pm}\partial_{\pm}\Xi_{\pm\pm}+2\Xi_{\pm\pm}\partial_{\pm}Y^{\pm}-\frac{1}{2}\partial^3_{\pm}Y^{\pm}.
\end{align}
The conformal factor $\varphi$ transforms only under $\omega$ while $\Xi_{\pm\pm}$ transform as the components of an anomalous $2$-dimensional CFT energy-momentum tensor \cite{Polchinski:1998rq}. Thanks to the redefinition of the residual diffeomorphisms generators \eqref{24.1} we have isolated the total part of the asymptotic symmetries that induces a Weyl rescaling of the boundary metric. This is in agreement with what is found in Appendix \ref{AppB}, where it is shown, using Chern--Simons formulation, that $\varphi$ completely decouples from the remaining dynamical content of the theory. Another more straightforward way to introduce $\omega$ is to require that the residual vector fields of \eqref{3} induce asymptotically a Weyl rescaling of the boundary metric
\begin{align}
\label{24.3}
\mathcal{L}_{\un\xi}\gamma_{ab}= 2\omega\rho^{-2}g_{ab}^{(0)}+O(\rho^0).
\end{align}
This equation leads to
\begin{align}
\label{24.4}
D^{(0)}_a Y_b+D^{(0)}_bY_b=2(\omega+\sigma)g^{(0)}_{ab}, 
\end{align}
which implies \eqref{24.1}.

Note that from the definition \eqref{T} of the holographic stress-energy tensor and from \eqref{14} it follows that, under a residual Weyl transformation, $T_{ab}$ transforms as
\begin{align}
\label{trT}
\delta_{(\omega,0)}T_{ab}=\frac{c}{12\pi}(D^{(0)}_aD^{(0)}_b\omega-g^{(0)}_{ab}\square{}^{(0)}\omega).
\end{align}
Hence, if we were to require that the vector field generating Weyl transformations satisfied
\begin{align}
\label{trTtr}
\delta_{(\omega,0)}T_{a}{}^a=-2\omega T_{a}{}^a-\frac{c}{12\pi}\square{}^{(0)}\omega\equiv-2\omega T_{a}{}^a ,
\end{align}
then the trace of $T_{ab}$, or equivalently $R^{(0)}$, would transform as a Weyl scalar of weight $-2$. This condition automatically implies that $\omega$ is an harmonic function
\begin{align}
\label{cond}
\square\omega=0,
\end{align}
whose general solution is
\begin{align}
\label{om}
\omega(x^+,x^-)=\omega^+(x^+)+\omega^-(x^-).
\end{align}
In the following, we will refer to this situation as the $\omega$-chiral case. Note that requiring the gauge parameter $\omega$ to satisfy \eqref{om} implies that $\varphi$ can vary under the action of the asymptotic symmetry group only as
\begin{align}
\label{varf}
\delta_{(\omega,Y^{\pm})}\varphi=\omega^+(x^+)+\omega^-(x^-).
\end{align}
We will return to the interpretation of \eqref{cond} in Section \ref{sec4}. For the moment let us just note that, under \eqref{cond}, even if the solution space does not admit a chiral splitting, its variation $\delta_{\un\xi}\chi$ can be decomposed into two sectors with definite chiralities, $\delta_{\un\xi^{\pm}}\chi=\{\delta_{(\omega^{\pm},Y^{\pm})}\Xi_{\pm\pm},\delta_{(\omega^{\pm},Y^{\pm})}\varphi\}$.
\subsection{Asymptotic Symmetry Algebra}
\label{SS7}

The on-shell residual diffeomorphisms generator in light-cone coordinates is
\begin{align}
\label{25}
\xi^{\rho}=\rho\sigma(x),\qquad \xi^{\pm}=Y^{\pm}(x^{\pm})+\rho^2\ell^2e^{-2\varphi}\partial_{\mp}\sigma+\rho^4\ell^2e^{-4\varphi}\left[\partial_{\mp}\sigma g^{(2)}_{+-}+\partial_{\pm}\sigma g^{(2)}_{\pm\pm}\right]+\mathcal{O}(\rho^6),
\end{align}
whereas the algebra is
\beqn
\label{27}
&\big[\xi_1,\xi_2\big]_{M}^{\rho}=\hat{\xi}^{\rho}=\rho\hat{\sigma},\qquad  \hat{\sigma}=Y_1^{+}\partial_{+}\sigma_2+Y_1^{-}\partial_{-}\sigma_2-Y_2^{+}\partial_{+}\sigma_1-Y_2^{-}\partial_{-}\sigma_1,&\\
\label{28}
& \big[\xi_1,\xi_2\big]_{M}^{\pm}=\hat{\xi}^{\pm}=\hat{Y}^{\pm}+\rho^2\ell^2e^{-2\varphi}\partial_{\mp}\hat{\sigma}+\mathcal{O}(\rho^4),\qquad \hat{Y}^{\pm}=Y_1^{\pm}\partial_{\pm}Y_2^{\pm}-Y_2^{\pm}\partial_{\pm}Y_1^{\pm}.&
\eeqn
This algebra is a semi-direct sum: by denoting an element of the algebra as the pair $(\sigma,Y^\pm)$, the Lie bracket between two elements is $\big[(\sigma_1,Y_1^\pm),(\sigma_2,Y_2^\pm)\big]=(\hat{\sigma},\hat{Y}^\pm)$, where $\hat{\sigma}$ and $\hat{Y}^{\pm}$ are given in \eqref{27} and \eqref{28}.

We now reformulate the algebra in terms of the parameter $\omega$ introduced in \eqref{24.1}. The on-shell generator is
\beqn
\label{25.1}
&&\xi^{\rho}=\rho\left(\Omega_{\un Y}-\omega\right),\\
&& \xi^{\pm}=Y^{\pm}+\rho^2\ell^2e^{-2\varphi}\partial_{\mp}\left(\Omega_{\un Y}-\omega\right)+\rho^4\ell^2e^{-4\varphi}\left[\partial_{\mp}\left(\Omega_{\un Y}-\omega\right)g^{(2)}_{+-}+\partial_{\pm}\left(\Omega_{\un Y}-\omega\right)g^{(2)}_{\pm\pm}\right]+\mathcal{O}(\rho^6).
\eeqn
Notice that this reformulation introduces a field dependence in $\xi^\rho$, which was previously absent. This implies that we need to use the modified Lie bracket also for this component. We now obtain
\begin{align}
\label{27.1}
&\big[\xi_1,\xi_2\big]_{M}^{\rho}=\hat{\xi}^{\rho}=\rho\left(\Omega_{\hat{\un Y}}-\hat{\omega}\right),\qquad \hat{Y}^{\pm}=Y_1^{\pm}\partial_{\pm}Y_2^{\pm}-Y_2^{\pm}\partial_{\pm}Y_1^{\pm},\qquad \hat{\omega}=0,
\end{align}
and, as before, 
\begin{align}
\label{28.1}
&\partial_{\rho}\left(\big[\xi_1,\xi_2\big]^{\pm}_M\right)=-\frac{\ell^2}{\rho^2}g^{ab}\partial_{b}\left(\big[\xi_1,\xi_2\big]^{\rho}_M\right),\qquad\lim_{\rho\to 0}\left(\big[\xi_1,\xi_2\big]^{\pm}_M\right)=\hat{Y}^{\pm}.
\end{align}
Integrating these equations leads to 
\beq
\big[\xi_1,\xi_2\big]^{\pm}_M=\hat{\xi}^{\pm}=\hat{Y}^{\pm}+\rho^2\ell^2e^{-2\varphi}\partial_{\mp}\left(\Omega_{\hat{\un Y}}-\hat{\omega}\right)+\rho^4\ell^2e^{-4\varphi}\left[\partial_{\mp}\left(\Omega_{\hat{\un Y}}-\hat{\omega}\right)g^{(2)}_{+-}+\partial_{\pm}\left(\Omega_{\hat{\un Y}}-\hat{\omega}\right)g^{(2)}_{\pm\pm}\right]+\mathcal{O}(\rho^6),\label{Ypm}
\eeq
where $\hat{Y}^\pm$ and $\hat{\omega}$ are defined in \eqref{27.1}.

With this set of independent generators of variations in solution space $(\omega,Y^\pm)$, the asymptotic symmetry algebra is thus a direct sum of two copies of the Witt algebra with the abelian ideal of Weyl rescalings. In order to describe the asymptotic symmetry algebra in terms of a basis, we use the notation established in \eqref{AKV} with the subscript $(\omega,Y^\pm)$ such that
\begin{align}
\label{31.0}
\un\xi_{(\omega,Y^\pm)}=\xi^{\rho}_{(\omega,Y^\pm)}\partial_{\rho}+\xi^{+}_{(\omega,Y^\pm)}\partial_{+}+\xi^{-}_{(\omega,Y^\pm)}\partial_{-}.
\end{align}
Consider the vector field $\un\xi_{Y^\pm}:= \un\xi_{(0,Y^\pm)}=\xi^{\rho}_{Y^\pm}\partial_{\rho}+\xi^{+}_{Y^\pm}\partial_{+}+\xi^{-}_{Y^\pm}\partial_{-}$ and the mode expansions $Y_1^{\pm}\sim e^{inx^{\pm}}$ and $Y_2^{\pm}\sim e^{imx^{\pm}}$. Computing the mode decomposition of $\hat{Y}^{\pm}$ in \eqref{27.1}
\begin{align}
\label{31.3}
\hat{Y}^{+}=i(n-m)e^{i(n+m)x^+},\hspace{1cm}\hat{Y}^{-}=i(n-m)e^{i(n+m)x^-},
\end{align}
we gather
\begin{align}
\label{31.4}
\big[\un \xi^\pm_n,\un \xi^\pm_m\big]_{M}=i(n-m)\un \xi^\pm_{n+m}, \qquad \big[\un \xi^\pm_n,\un \xi^\mp_m\big]_{M}=0,
\end{align}
where we replaced the $Y^\pm$ subscript by the mode number $\un\xi_{Y^\pm}\mapsto \un\xi^\pm_{n}$.\footnote{Here the notation $\un\xi^\pm_{n}$ means that $\un\xi^+_{n}$ is the the full vector $\un\xi_{Y^+}$ with $Y^+\sim e^{inx^+}$ and $Y^-=0$, while $\un\xi^-_{n}$ is the vector $\un\xi_{Y^-}$ with $Y^+=0$ and $Y^-\sim e^{inx^-}$.} We thus have two copies of the Witt algebra, which is expected since for $\omega=0$ we reach BH boundary conditions, where this algebra has already been derived, \cite{Brown:1986nw}.

Set now $Y^\pm=0$ and consider $\un\zeta_{\omega}:= \un\xi_{(\omega,0)}=\zeta^{\rho}_{\omega}\partial_{\rho}+\zeta^{+}_{\omega}\partial_{+}+\zeta^{-}_{\omega}\partial_{-}$. Expanding $\omega_1\sim e^{ipx^+}e^{iqx^-}$ and $\omega_2\sim e^{irx^+}e^{isx^-}$ the algebra reads
\begin{align}
\label{31.5}
\big[\un\zeta_{pq},\un\zeta_{rs}\big]_{M}=0, \qquad 
\big[\un\xi^\pm_n,\un\zeta_{rs}\big]_{M}=0.
\end{align}
where we replaced the $\omega$ subscript by the mode numbers $\un\zeta_{\omega}\mapsto \un\zeta_{pq}$. Denoting an element of the algebra as the pair $(\omega,Y^\pm)$, the Lie bracket between two elements is $\big[(\omega_1,Y^\pm_1),(\omega_2,Y^\pm_2)\big]=(0,\hat{Y}^\pm)$.

 In the particular subclass of $\omega$ satisfying \eqref{om}, i.e. the $\omega$-chiral case, we can consider the algebra of the left and right Weyl sectors separately. Expanding $\omega^{\pm}\sim e^{ipx^{\pm}}$ we denote by $\un\zeta^{+}_{p}$ the vector $\un\zeta_{\omega}$ with $\omega^+\sim e^{ipx^+}$ and $\omega^-=0$ and by $\un\zeta^{-}_{p}$ the vector $\un\zeta_{\omega}$ with $\omega^-\sim e^{ipx^-}$ and $\omega^+=0$. The algebra now becomes
\begin{align}
\label{31.56}
\big[\un\zeta_p^{\pm},\un\zeta_{q}^{\pm}\big]_M=0,\qquad \big[\un\zeta_p^{\pm},\un\zeta_{q}^{\mp}\big]_M=0, \qquad \big[\un\xi^{\pm}_n,\un\zeta^{\pm}_p\big]_M=0,\qquad \big[\un\xi^{\pm}_n,\un\zeta^{\mp}_p\big]_M=0.
\end{align}

\section{Charges and Algebra}
\label{S2}
This Section is devoted to the study of asymptotic charges under the boundary conditions spelled above. We will discuss the charge algebra: we retrieve the usual Virasoro double copy,  plus a Weyl sector with a non-trivial central extension.
\subsection{Surface Charges}
\label{S3}
Surface charges are computed using the prescription given in \cite{Barnich:2001jy}
\begin{align}
\label{32}
\cancel{\delta}Q_{\un\xi}[h,g]=\frac{1}{16\pi G}\int_{S^1_{\infty}} \frac{1}{2}\varepsilon_{\mu\nu\alpha}\D x^{\alpha}K_{\un\xi}^{\mu\nu}[g,h]=\frac{1}{16\pi G}\int_0^{2\pi}\D \phi K_{\un\xi}^{\rho t}[g,h].
\end{align}
Here $h_{\mu\nu}=\delta g_{\mu\nu}$ are the on-shell variations of the metric, the integration is on the circle at infinity spanned by $\phi$ and we use the convention $\varepsilon_{\rho t\phi}=1$. The antisymmetric tensor $K^{\mu\nu}_{\un\xi}[g,h]$ in \eqref{32} is explicitly given by
\begin{align}
\label{33}
&K^{\mu\nu}_{\un\xi}[g,h]=\sqrt{-g}\big[\xi^{\nu}D^{\mu}h-\xi^{\nu}D_{\sigma}h^{\mu\sigma}+\xi_{\sigma}D^{\nu}h^{\mu\sigma}+\frac{1}{2}hD^{\nu}\xi^{\mu}+\frac{1}{2}h^{\nu\sigma}(D^{\mu}\xi_{\sigma}-D_{\sigma}\xi^{\mu})-(\mu\leftrightarrow\nu)\big].
\end{align}
Charges are computed at fixed time $t$ and radial coordinate $\rho$ approaching the boundary and directly in the $\omega$ parametrization. The charges associated to $Y^\pm$ with $\omega=0$ are (as for the vector fields we define $\cancel{\delta} Q_{Y^\pm}[g,h]:=\cancel{\delta} Q_{(0,Y^\pm)}[g,h]$)
\begin{align}
\label{41}
\cancel{\delta}Q_{Y^\pm}[g,h]=\frac{\ell}{8\pi G}\int_0^{2\pi}\D\phi\left(Y^-\delta\Xi_{--}+Y^+\delta\Xi_{++}\right).
\end{align}
To obtain these charges we used $\partial_{\phi}=\partial_+-\partial_-$ and integrated out total $\phi$ derivative terms. The $Y^\pm$ charges are thus integrable:
\begin{align}
\label{42}
Q_{Y^\pm}[g]=\frac{\ell}{8\pi G}\int_{0}^{2\pi}\D\phi\left(Y^+\Xi_{++}+Y^-\Xi_{--}\right).
\end{align}
These are the usual conserved charges, found also with BH boundary conditions. The $Q_{Y^{\pm}}[g]$ in \eqref{42} are computed with respect to the background metric $\bar g$ defined by $\Xi_{\pm\pm}=0$, which is the BTZ black hole with vanishing mass and angular momentum. The ones computed with respect to the global AdS${}_3$ background can be obtained shifting $\Xi_{\pm\pm}\rightarrow \Xi_{\pm\pm}+\frac{1}{4}$ in \eqref{42}, \cite{Barnich:2012aw}. 

The Weyl sector, found by setting $Y^\pm=0$, is also integrable and given by
\begin{align}
\label{43}
Q_{\omega}[g]=\frac{\ell^2}{8\pi G}\int_0^{2\pi}\D\phi(\varphi\partial_t\omega-\omega\partial_t\varphi).
\end{align}
The same charges can be obtained using the Iyer--Wald prescription \cite{Wald:1993nt,Iyer:1994ys}. While these are the most general Weyl charges in our setup, we now restrict attention to the case \eqref{om}, i.e. $\omega=\omega^++\omega^-$. Correspondingly, the Weyl charges decompose as
\begin{align}
\label{Wc}
Q_{\omega}[g]=-\frac{\ell}{4\pi G}\int_{0}^{2\pi}\D\phi\left(\omega^+\partial_+\varphi+\omega^-\partial_-\varphi\right)\equiv Q_{\omega^+}[g]+Q_{\omega^-}[g],
\end{align}
where we have integrated by parts. Note that these charges split into two pieces, generating the  chiral and anti-chiral transformations of $\varphi$.
Contrarily to the $Y^\pm$ sector, $Q_{\omega}$ is not conserved. This is due to the presence of a non-vanishing symplectic flux through the boundary, as we will emphasize in Section \ref{sec4}.

We would like to stress that the main result of our paper is the computation of the surface charges including a new non-trivial Weyl sector. These additional interesting charges are finite, integrable but not conserved. These features make them special. Other examples of non-conserved integrable charges at finite boundaries are discussed in \cite{Adami:2020ugu}. We now proceed to compute their algebra.
 
\subsection{Charge Algebra} 
\label{SS9}
We now verify that the surface charges, under the Poisson brackets, form a projective representation  of the asymptotic symmetry algebra with modified Lie brackets
\begin{align}
\label{44}
\big\{Q_{\un\xi_1}[g],Q_{\un\xi_2}[g]\big\}=\delta_{\un \xi_2}Q_{\un\xi_1}[g]=Q_{\big[\un\xi_1,\un\xi_2\big]_M}[g]+\mathcal{K}_{\un\xi_1,\un\xi_2},
\end{align}
where $\mathcal{K}_{\un\xi_1,\un\xi_2}$ is the central extension. 

We start by computing $\delta_{\un\xi_2}Q_{\un\xi_1}[g]$. Defining the integrand $K^{\rho t}_{\un\xi}[g]$ as $K^{\rho t}_{\un\xi}[g,h]=\delta K^{\rho t}_{\un\xi}[g]$, we have
\begin{align}
\label{45}
\delta_{(\omega_2,Y_2^\pm)}K^{\rho t}_{(\omega_1,Y_1^\pm)}[g]=\frac{\ell}{8\pi G}\left(Y_1^+\delta_{(\omega_2,Y_2^\pm)}\Xi_{++}+Y_1^-\delta_{(\omega_2,Y_2^\pm)}\Xi_{--}+\ell\delta_{(\omega_2,Y_2^\pm)}\varphi\partial_t\omega_1-\ell\omega_1\partial_t\delta_{(\omega_2,Y_2^\pm)}\varphi\right).
\end{align}
We work separately for the $Y^\pm$ and $\omega$ parts. This can be done because $\Xi_{\pm\pm}$ and $\varphi$ transform independently under $Y^\pm$ and $\omega$, respectively. For the $Y^\pm$ sector, after a straightforward computation, this yields the well-known BH central extension \cite{Brown:1986nw}
\begin{align}
\label{49}
\mathcal{K}_{\un\xi_{Y_1},\un\xi_{Y_2}}=\frac{1}{8\pi G}\int_{0}^{2\pi}\D \phi\left(\partial_{\phi}Y_1^t\partial^2_{\phi}Y_2^{\phi}-\partial_{\phi}Y_2^t\partial^2_{\phi}Y_1^{\phi}\right).
\end{align}

Consider now the Weyl sector, obtained by setting $Y^\pm=0$ with non-vanishing $\omega$. We have
\begin{align}
\label{50}
\delta_{(\omega_2,0)} K^{\rho t}_{(\omega_1,0)}[g]=\frac{\ell^2}{8\pi G}\left(\omega_2\partial_t\omega_1-\omega_1\partial_t\omega_2\right).
\end{align}
Here, since the asymptotic symmetry algebra is  abelian we have $Q_{\big[\un\zeta_1,\un\zeta_2\big]_M}[g]=0$. Thence
\beq
\label{WCEx}
\mathcal{K}_{\un\zeta_1,\un\zeta_2}=\frac{\ell^2}{8\pi G}\int_{0}^{2\pi}\D \phi \left(\omega_2\partial_t\omega_1-\omega_1\partial_t\omega_2\right).
\eeq
The complete charge algebra is
$\big\{Q_{\un\xi_1}[g],Q_{\un\xi_2}[g]\big\}=\delta_{\un\xi_2}Q_{\un\xi_1}[g]=Q_{\hat{\un\xi}}[g]+\mathcal{K}_{\un\xi_1,\un\xi_2}$,
where $\hat{\un\xi}=\big[\un\xi_1,\un\xi_2\big]_M$ are gathered in \eqref{27.1} and \eqref{Ypm}. The total central extension is 
$\mathcal{K}_{\un\xi_1,\un\xi_2}=\mathcal{K}_{\un\xi_{Y_1},\un\xi_{Y_2}}+\mathcal{K}_{\un\zeta_1,\un\zeta_2}$
To this expression contributes the ordinary BH central charge plus an additional term coming from the Weyl rescalings of the boundary metric. 
The $Y^\pm$ sector of the central extension evaluated on the vector fields mode decomposition $\un\xi^\pm_n$ and $\un\xi^\pm_m$ (modes of $\un\xi_{(0,Y^\pm)}$) is
\begin{align}
\label{52.25}
\mathcal{K}_{\un\xi^\pm_n,\un\xi^\pm_m}=-im^3\frac{c^\pm}{12}\delta_{n+m,0},\qquad \mathcal{K}_{\un\xi^\pm_n,\un\xi^\mp_m}=0,\qquad c^{\pm}=c=\frac{3\ell}{2 G}.
\end{align}
On the other hand, the central extension for the modes decomposition of the Weyl sector yields
\begin{align}
\label{52.31}
\mathcal{K}_{\un\zeta_{pq},\un\zeta_{rs}}=-i(r-q)c_{_W}\omega_{q+s,q+s}\delta_{p+r,q+s}, \qquad c_{_W}=\frac{\ell}{2 G}.
\end{align}
The total charge algebra then reads
\begin{align}
\label{A10}
&\big\{Q_{\un\xi_n^\pm}[g],Q_{\un\xi_m^{\pm}}[g]\big\}=i(n-m)Q_{\un\xi^{\pm}_{n+m}}[g]-im^3\frac{c^{\pm}}{12}\delta_{n+m,0},	\\
&\big\{Q_{\un\xi_n^\pm}[g],Q_{\un\xi_m^{\mp}}[g]\big\}=0,	\\
\label{mix}
&\big\{Q_{\un\zeta_{pq}}[g],Q_{\un\zeta_{rs}}[g]\big\}=-i(r-q)c_{_W}e^{2i(q+s)t/\ell}\delta_{p+r,q+s},	\\
&\big\{Q_{\un\xi_n^{\pm}}[g],Q_{\un\zeta_{pq}}[g]\big\}=0,
\end{align}
This algebra is the direct sum of two centrally extended Virasoro sectors and the centrally extended Weyl sector. We note that the Weyl central extension is explicitly time dependent. As such, we are dealing here with a one-parameter family of algebras, labelled by the time slice $t$ at which the charges are computed.

The total expression $\mathcal{K}_{\un\xi_1,\un\xi_2}$ is indeed a $2$-cocycle because it satisfies
\begin{align}
\label{52.1}
\mathcal{K}_{[\un\xi_1,\un\xi_2]_M,\un\xi_3}+\mathcal{K}_{[\un\xi_3,\un\xi_1]_M,\un\xi_2}+\mathcal{K}_{[\un\xi_2,\un\xi_3]_M,\un\xi_1}=0.
\end{align}
This equation is automatically satisfied for the Weyl sector and the mixed sector, while in the Witt sectors it is proved as usual. Furthermore, since the Virasoro central extension is non-trivial and any $2$-cocycles of an Abelian algebra cannot be a coboundary, \eqref{WCEx} is non-trivial.

Again, in the $\omega$-chiral case, the central extension for the Weyl left- and right-movers simplifies to
\begin{align}
\label{52.3}
\mathcal{K}_{\un\zeta^{\pm}_{p},\un\zeta^{\pm}_{q}}=ipc_{_W}^{\pm}\delta_{p+q,0},\qquad\mathcal{K}_{\un\zeta^{\pm}_{p},\un\zeta^{\mp}_{r}}=0
\end{align}
The total charge algebra then reads 
\begin{align}
\label{chargealg1}
&\big\{Q_{\un\xi_n^\pm}[g],Q_{\un\xi_m^{\pm}}[g]\big\}=i(n-m)Q_{\un\xi^{\pm}_{n+m}}[g]-im^3\frac{c^{\pm}}{12}\delta_{n+m,0},	\\
\label{chargealg2}
&\big\{Q_{\un\xi_n^\pm}[g],Q_{\un\xi_m^{\mp}}[g]\big\}=0,	\\
\label{chargealg3}
&\big\{Q_{\un\zeta_p^{\pm}}[g],Q_{\un\zeta_q^{\pm}}[g]\big\}=ipc^{\pm}_{_W}\delta_{p+q,0},	\\
\label{chargealg4}
&\big\{Q_{\un\zeta_p^{\pm}}[g],Q_{\un\zeta_q^{\mp}}[g]\big\}=0,\\
\label{chargealg5}
&\big\{Q_{\un\xi_n^{\pm}}[g],Q_{\un\zeta_p^{\pm}}[g]\big\}=0,\\
\label{chargealg6}
&\big\{Q_{\un\xi_n^{\pm}}[g],Q_{\un\zeta_p^{\mp}}[g]\big\}=0.
\end{align}
In this particular framework the Weyl central extension does not depend on time and therefore the one-parameter family of algebras reduces to a Kac-Moody current algebra. The algebra \eqref{chargealg1}-\eqref{chargealg6}, up to redefinition of generators, is the same as the one found in \cite{Troessaert:2013fma}, as reviewed in Appendix \ref{AppA}.
\section{Holographic Aspects}\label{sec4}
Thanks to the AdS/CFT dictionary \cite{tHooft:1993dmi, Susskind:1994vu, Maldacena:1997re, Gubser:1998bc, Witten:1998qj}, we know that the bulk gravity theory is dual to a boundary field theory. As long as the former is in the classical limit, the latter is strongly coupled. Therefore, little is known about it: we cannot construct its perturbative action but we still have access to non-perturbative features such as quantum symmetries expressed in terms of Ward--Takahashi identities of the path integral partition function \cite{Ward:1950xp, Takahashi:1957xn}. The goal of this Section is to show that there is a breaking in the conservation law of the Weyl current, which has a holographic dual counterpart as a boundary anomalous Ward--Takahashi identity \cite{Bilal:2008qx}. Before proceeding let us briefly review the emergence of the Weyl anomaly in the context of holographic renormalization.

The renormalized action for General Relativity in asymptotically locally AdS${}_3$ spacetimes is defined \cite{Henningson:1998gx,Skenderis:2002wp,Compere:2008us} as $S[g]=\lim_{\epsilon\to 0}S_{\epsilon}[g]$ where $S_{\epsilon}[g]$ is the regularized action, given by \\
\begin{align}
\label{renAc}
S_{\epsilon}[g]=\frac{1}{16\pi G}\int_{M_{\epsilon}}\D^3x\sqrt{-g} \left(R-\frac{2}{\ell^2}\right)+\frac{1}{16\pi G}\int_{\partial M_{\epsilon}}\D^2x \sqrt{-\gamma}\left(2K-\frac{2}{\ell}+\frac{\ell}{4}R^{(0)}\log\epsilon\right),
\end{align}\\
where $K$ is the trace of the extrinsic curvature of the constant $\rho$ hypersurface and the last two terms are the standards counterterms. The renormalized action $S[g]$ is therefore obtained by first introducing a cut-off at $\rho=\epsilon$ that allows the divergences to cancel and then by setting the limit $\epsilon\rightarrow 0$. Taking an on-shell variation of $S[g]$ yields\footnote{For the Chern-Simons reformulation of the variational problem, see Appendix \ref{AppB}.}
\begin{align}
\label{var}
\delta S[g]=	\frac{1}{2}\int_{\partial M}\D^2x\sqrt{-g^{(0)}}T^{ab}\delta g^{(0)}_{ab}=\frac{c}{24\pi}\int_{\partial M}\D^2x\sqrt{-g^{(0)}}R^{(0)}\delta\varphi,
\end{align}
where in the last step we have used the conformally flat parametrization. Hence, with our choice of boundary conditions, the variational problem is not well-defined \cite{Papadimitriou:2005ii,Compere:2008us}.
Specifying $\delta$ to be the variation \eqref{24.2} induced by a Weyl diffeomorphism so that $\delta_{\omega}g^{(0)}_{ab}=2\omega g^{(0)}_{ab}$, we get
\begin{align}
\label{W An}
\delta_{\omega} S[g]=\frac{c}{24\pi}\int_{\partial M}\D^2x\sqrt{-g^{(0)}} R^{(0)}\omega\equiv \int_{\partial M}\D^2x \sqrt{-g^{(0)}} \mathcal{A}\, \omega,\qquad \mathcal{A}=\frac{c}{24\pi} R^{(0)},
\end{align}
which is the standard expression for the Weyl anomaly in AlAdS${}_3$ spacetimes. Note that we define $\mathcal{A}$ to be the integrand coefficient of $\omega$ in $\delta_{\omega}S[g]$ \cite{Deser:1993yx,Henningson:1998gx}.
Taking a variation of \eqref{var} yields the induced symplectic structure on the boundary $\partial M$ \cite{Crnkovic:1986ex,Compere:2018aar,Alessio:2019cae}
\begin{align}
\label{53}
\pmb{\omega}(\delta_1,\delta_2)=	\frac{1}{2}\int_{\partial M} \D^2x\delta_1 \left(\sqrt{-g^{(0)}}T^{ab}\right)\wedge\delta_2 g^{(0)}_{ab}=-\frac{1}{8\pi G}\int_0^{2\pi}\D\phi\int_{t_1}^{t_2} \D t\left(\square\delta_1\varphi\wedge\delta_2\varphi\right),
\end{align}
where we have used the conformally flat parametrization and where the flat laplacian is defined as $\square=-\ell^2\partial^2_t+\partial^2_{\phi}$. Integrating by parts in $t$ we have, discarding total $\phi$ derivatives, that 
\begin{align}
\label{55}
\pmb{\omega}(\delta_1,\delta_2)=-\frac{\ell^2}{8\pi G}\int_0^{2\pi}\D\phi\int_{t_1}^{t_2}\D t\left[\partial_t\left(\delta_1\varphi\wedge\partial_t\delta_2\varphi\right)\right]=-\frac{\ell^2}{8\pi G}\int_0^{2\pi}\D\phi\left[\delta_1\varphi\wedge\partial_t\delta_2\varphi\right]^{{}^{t_2}}_{{}_{t_1}}.
\end{align}

This shows that the difference in time of the Weyl charges is equal to the symplectic flux contracted with a Weyl generating vector field,\footnote{This result can be derived in first order formalism, \cite{Barnich:2019vzx}.}
\begin{align}
\label{56}
\pmb{\omega}(\delta_{\omega},\delta)=\frac{\ell^2}{8\pi G}	\delta\int_0^{2\pi}\D\phi\left[\varphi\partial_t\omega-\omega\partial_t\varphi\right]^{{}^{t_2}}_{{}_{t_1}}=\delta Q_{\omega}(t_2)-\delta Q_{\omega}(t_1).
\end{align}
Therefore the Weyl charges are not conserved but integrable, as mentioned above. We proceed now to reduce the theory to the $\omega$-chiral case.

We start by noticing that, taking a Weyl variation of equation \eqref{W An}, we obtain
\begin{align}
\label{varWan}
\delta_{\omega_1} \delta_{\omega_2}S[g]=-\frac{1}{8 G}\int_{\partial M}\D^2x \omega_1 \square\omega_2.
\end{align}
We thus see that the effect of constraining the form of the asymptotic symmetry generators according to \eqref{cond} is to make $\delta_{\omega_1} \delta_{\omega_2}S[g]$ vanishing. In other words, \eqref{cond} means that we are not allowing the integrated Weyl anomaly to vary under the action of the asymptotic symmetry algebra.
This means that, under the above-spelled condition on $\omega(x)$, the operator $\delta_{\omega}$ on the functional $S[g]$ is a cocycle $\delta_{\omega_1}\delta_{\omega_2}S[g]=0$ \cite{Bonora:1985cq,Mazur:2001aa,Boulanger:2007ab,Boulanger:2007st}. From now on we will impose $\square\omega=0$ and comment on some holographic aspects in this framework.

We now proceed to construct a Weyl current \cite{Barnich:2013axa}. This procedure is well-known for the Virasoro sector, where the currents combine in the stress tensor of the boundary dual theory, and its conservation is interpreted in the bulk as Einstein equations, while in the boundary as the Ward--Takahashi identity for the transformations generated by $\xi^a_{(0,Y^{\pm})}$. In a similar fashion, given that the condition $\square \omega=0$ ensures we deal with chiral and anti-chiral Weyl charges generators \eqref{Wc}, we can define two Weyl currents. Starting from \eqref{33} we introduce
\beq
K_{(\omega,0)}^{\rho a}[g]=K_{(\omega^+,0)}^{\rho a}[g]+K_{(\omega^-,0)}^{\rho a}[g].
\eeq
Before giving their explicit expression, we first use the ambiguity in defining $K_{\un\xi}^{\mu\nu}[g,h]$
\beqn
\tilde K^{\rho a}_{(\omega^+,0)}[g] &=& K^{\rho a}_{(\omega^+,0)}[g]+\pa_b F^{[ba]}_{\omega^+},\\
\tilde K^{\rho a}_{(\omega^-,0)}[g] &=& K^{\rho a}_{(\omega^-,0)}[g]+\pa_b F^{[ba]}_{\omega^-}.
\eeqn
Choosing
\beq
F_{\omega^+}^{+-} =- {\ell\over 8\pi G}\varphi \omega^+,\qquad
F_{\omega^-}^{+-} ={\ell\over 8\pi G}\varphi \omega^-,
\eeq
we obtain
\beq
\tilde K^{\rho +}_{(\omega^+,0)}[g] = 0,\qquad 
\tilde K^{\rho -}_{(\omega^+,0)}[g] = -{\ell \over 4\pi G}\omega^+\pa_+ \varphi,
\eeq
and
\beq
\tilde K^{\rho +}_{(\omega^-,0)}[g] =-{ \ell \over 4 \pi G}\omega^- \pa_- \varphi ,\qquad
\tilde K^{\rho -}_{(\omega^-,0)}[g] = 0.
\eeq
These are the integrands of the chiral and anti-chiral Weyl charges found in \eqref{Wc}. We can now introduce the two Weyl currents $\un{\tilde{J}}_{\omega^+}=\tilde J^a_{\omega^+}\pa_a=\tilde J^+_{\omega^+}\pa_++\tilde J^-_{\omega^+}\pa_-$ and $\un{\tilde{J}}_{\omega^-}=\tilde J^a_{\omega^-}\pa_a=\tilde J^+_{\omega^-}\pa_++\tilde  J^-_{\omega^-}\pa_-$ for the two chirality sectors as
\beqn
 \tilde K^{\rho a}_{(\omega^+,0)}[g] &=& \sqrt{-g^{(0)}}\omega^+ \tilde J^a_{\omega^+},\\
 \tilde K^{\rho a}_{(\omega^-,0)}[g] &=& \sqrt{-g^{(0)}}\omega^- \tilde J^a_{\omega^-},
\eeqn
such that the currents are tensors ($K_{\un\xi}^{\mu\nu}[g,h]$ in \eqref{33} is a tensor density) and they do not depend by the gauge parameters $\omega^+$ and $\omega^-$. Their explicit expressions, using $\D s^2_{\text{bdy}}=-e^{2\varphi} \D x^+\D x^-$, are
\beq
\tilde J^+_{\omega^+}= 0,\qquad 
\tilde J^-_{\omega^+}= -{\ell e^{-2\phi} \over 2\pi G}\pa_+ \varphi,
\eeq
and
\beq
\tilde J^+_{\omega^-}=-{ \ell e^{-2\phi} \over 2 \pi G}\pa_- \varphi ,\qquad 
\tilde J^-_{\omega^-}= 0.
\eeq
We eventually compute the boundary covariant divergence of these two currents and find:
\beqn
D_a^{(0)}\tilde J^a_{\omega^+}&=&-{\cal A} ,\\
D_a^{(0)}\tilde J^a_{\omega^-}&=&-{\cal A} ,
\eeqn
where ${\cal A}$ is the anomaly integrand coefficient defined in \eqref{W An}. We have thus shown that the Weyl currents are not conserved due to the presence of the anomaly \cite{Bilal:2008qx}. The boundary Weyl symmetry is broken, for the bulk counterpart Weyl charges are not conserved. This process is driven by the anomaly coefficient: for flat boundary metrics the current is conserved \cite{Troessaert:2013fma}, as we thoroughly review in Appendix \ref{AppA}.

\section{Conclusions}
\label{S6}
We summarize here our results and offer some possible outlooks of the present work. In the first part of the paper we have analyzed the asymptotic structure of 3-dimensional General Relativity for AlAdS${}_3$ spacetimes. In the spirit of keeping diffeomorphisms generating Weyl rescalings of the boundary metric disentangled from those generated by $Y^{a}$, we imposed a specific set of boundary conditions, namely the boundary metric being conformally flat, with only the conformal factor $\varphi$ free to vary within the solution space. Correspondingly, we have computed the asymptotic symmetry algebra of this setup. The boundary conditions adopted here do not lead to a well-defined variational principle. Nonetheless, we have found finite and integrable, although not conserved, surface charges associated to the bulk diffeomorphisms generating Weyl transformations. Integrability of the charges allows us to construct the charge algebra, which admits a new central extension in the Weyl sector.

Concerning the holographic interpretation, the AdS/CFT dictionary predicts that bulk asymptotic symmetries are dual to boundary global symmetries of a putative field theory. Although the holographic interpretation of the Weyl sector has been widely investigated, this has not been done explicitly in terms of asymptotic symmetries. That is, boundary currents built out of bulk asymptotic Weyl charges, such that their non-conservation results in the anomalous Ward--Takahashi identity, have not been previously constructed. In Section \ref{sec4} we filled this gap by explicitly deriving these currents in the $\omega$-chiral case. 
 
In this manuscript we have not addressed the holographic interpretation corresponding to the most general variation of the boundary metric (i.e. $\omega$ not satisfying \eqref{cond}), which is certainly worth exploring. In this regard, a different choice of gauge in the bulk may be more suited, e.g. \cite{Ciambelli:2019bzz}. In particular, this raises the question on how the Weyl charges explicitly depend on the gauge condition \cite{Ciambelli:2020eba,Ciambelli:2020ftk,Compere:2019bua,Compere:2020lrt}. Another outlook is the extension of this work to higher dimensions. Specifically, we expect to unravel similar patterns in even-boundary dimensions, wherease it would also be interesting to investigate Weyl charges in odd-boundary dimensions. Furthermore, a suitable flat limit \cite{Barnich:2012aw,Ciambelli:2018wre} of these results might be relevant for the flat holography program \cite{Arcioni:2003xx,Dappiaggi:2005ci,Chandrasekaran:2020wwn,Laddha:2020kvp} and the recent developments in celestial CFT \cite{Strominger:2017zoo ,Donnay:2020guq,Ball:2019atb}. Eventually, on the macroscopic side of holography, i.e., in the fluid/gravity correspondence, it would be interesting to study the role of these boundary conditions from the fluid perspective \cite{Campoleoni:2018ltl}.

\section*{Acknowledgements}

We thank Andrea Campoleoni, Geoffrey Comp\`ere, Stephane Detournay, Adrien Fiorucci, Rob Leigh and C\'eline Zwikel for insightful discussions.  The work of LC is supported by the ERC Advanced Grant ``High-Spin-Grav". The work of RR was supported by the FRIA (FNRS, Belgium) and the Austrian Science Fund (FWF), project P 32581-N. The work of PM is supported in part by the National Natural Science Foundation of China under Grant No. 11905156 and No. 11935009.

\appendix

\section{Chiral Splitting of the Conformal Factor}
\label{AppA}

This Appendix is devoted to the comparison of this paper with \cite{Troessaert:2013fma}. There, one requires an additional boundary condition, namely
\begin{align}
\label{A1}
\square^{(0)}\varphi=0.
\end{align}
This implies that the variational principle is well defined. Indeed, the solution of \eqref{A1} is, in light-cone coordinates,
\begin{align}
\label{A2}
\varphi(x^+,x^-)=\varphi^+(x^+)+\varphi^-(x^-).
\end{align}
The boundary line element is thus
\begin{align}
\label{A3}
\D s^2_{\text{bdy}}=g^{(0)}_{ab}\D x^a\D x^b=-e^{2\varphi^+(x^+)}\D x^+e^{2\varphi^-(x^-)}\D x^-.
\end{align}
Notice in particular that, with these boundary conditions, the boundary metric is flat
\begin{align}
\label{A5}
R^{(0)}=8e^{-2\varphi}\pa_+\pa_-\varphi=0.
\end{align}
Clearly, in order to preserve \eqref{A1}, the parameter $\omega$ generating Weyl transformations must be of the form
\begin{align}
\label{A6}
\omega(x^+,x^-)=\omega^+(x^+)+\omega^-(x^-),
\end{align}
i.e. it admits a splitting into a chiral and an anti-chiral part. Thus, we can repeat the same arguments of Section \ref{S2} and the asymptotic symmetry algebra sector involving Weyl generators is given again by \eqref{31.56},
\begin{align}
\big[\un\zeta_p^{\pm},\un\zeta_{q}^{\pm}\big]_M=0,\qquad \big[\un\zeta_p^{\pm},\un\zeta_{q}^{\mp}\big]_M=0, \qquad \big[\un\xi^{\pm}_n,\un\zeta^{\pm}_p\big]_M=0,\qquad \big[\un\xi^{\pm}_n,\un\zeta^{\mp}_p\big]_M=0.
\end{align}
In this setup the Weyl charges become explictly
\begin{align}
\label{A7}
Q_{\omega}[g]=-\frac{\ell}{4\pi G}\int_0^{2\pi}\D\phi(\omega^+\partial_+\varphi^++\omega^-\partial_-\varphi^-)\equiv Q_{\omega^+}[g]+Q_{\omega^-}[g].
\end{align}
Furthermore, since the Weyl central extension \eqref{WCEx} is independent of $\varphi$, it is given again by 
\begin{align}
\label{A9}
{\mathcal K}_{\un\zeta_{p}^{\pm},\un\zeta_{q}^{\pm}}=ipc_{_W}^{\pm}\delta_{q+p,0},\qquad {\mathcal K}_{\un\zeta_{p}^{\pm},\un\zeta_{q}^{\mp}}=0,
\end{align}
just as in \eqref{52.3}. Therefore the centrally extended charge algebra with these boundary conditions is the same as \eqref{chargealg1}-\eqref{chargealg6}, 
\begin{align}
\label{A10}
&\big\{Q_{\un\xi_n^\pm}[g,h],Q_{\un\xi_m^{\pm}}[g]\big\}=i(n-m)Q_{\un\xi^{\pm}_{n+m}}[g]-im^3\frac{c^{\pm}}{12}\delta_{n+m,0},	\\
&\big\{Q_{\un\xi_n^\pm}[g],Q_{\un\xi_m^{\mp}}[g]\big\}=0,	\\
&\big\{Q_{\un\zeta_p^{\pm}}[g],Q_{\un\zeta_q^{\pm}}[g]\big\}=ipc_{_W}^{\pm}\delta_{p+q,0},	\\
&\big\{Q_{\un\zeta_p^{\pm}}[g],Q_{\un\zeta_q^{\mp}}[g]\big\}=0,\\
&\big\{Q_{\un\xi_n^{\pm}}[g],Q_{\un\zeta_p^{\pm}}[g]\big\}=0,\\
&\big\{Q_{\un\xi_n^{\pm}}[g],Q_{\un\zeta_p^{\mp}}[g]\big\}=0.
\end{align} Several comments are in order here. First of all, we remark that the charges obtained here are conserved, integrable and finite. This is expected: the non-conservation of our charges was due to the non-flatness of the boundary metric. Secondly, the charge algebra is not any longer explicitly time dependent. Lastly, note that in our basis the algebra is a direct sum of the Virasoro and the Weyl piece, while in \cite{Troessaert:2013fma} the algebra was represented as a semi-direct sum. This is ultimately a consequence of our field-dependent redefinition \eqref{24.1}.
\section{Chern--Simons Formulation}\label{AppB}

We reformulate here our results in the Chern--Simons formulation. This has a twofold purpose: it allows on the one hand to compare our results with \cite{Barnich2014TheDT} while on the other hand to perform the Gauss decomposition which outlines the role played by the Weyl anomaly and the absence of propagating bulk degrees of freedom. In particular, we will show that the conformal factor decouples from the dynamical fields of the theory. This Appendix extends to our boundary conditions results obtained originally in 
 \cite{Coussaert:1995zp,Henneaux:1999ib} and further discussed in \cite{Barnich2014TheDT, Donnay:2016iyk}.

\subsection{Conventions and Solution Space}

Three-dimensional General Relativity with a negative cosmological constant can be described by a Chern--Simons theory for an $\mathfrak{so}(2,2)$ valued connection \cite{Witten:1988hc, Witten:2007kt}. In particular, since $\mathfrak{so}(2,2)$ is isomorphic to $\mathfrak{sl}(2,\mathbb{R})\oplus\mathfrak{sl}(2,\mathbb{R})$,\footnote{We are going to refer to the two copies of $\mathfrak{sl}(2,\mathbb{R})$ as the left or chiral $\mathfrak{sl}(2,\mathbb{R})_L$ and right or anti-chiral $\mathfrak{sl}(2,\mathbb{R})_R$.} the Einstein--Hilbert action can be written, up to boundary terms, as the sum of two Chern--Simons actions 
\begin{align}
\label{FA1}
S_{EH}[A,\bar{A}]=S_{CS}[A]-S_{CS}[\bar{A}],
\end{align}
where we have denoted by $A$ and $\bar{A}$ the chiral and anti-chiral connections, respectively, and where
\begin{align}
S_{CS}[A]=-\kappa\int_{M}\D^3 x\text{Tr}\big(A\wedge dA+\frac{2}{3}A\wedge A\wedge A\big),\hspace{1cm}\kappa =\frac{\ell}{16\pi G}.
\end{align}
Following the conventions used in\cite{Barnich2014TheDT}, we choose the generators of  $\mathfrak{sl}(2,\mathbb{R})$ as
\begin{align}
\label{FA2}
j_{+}=-\frac{1}{\sqrt{2}}\left(\begin{matrix} 0 & 0 \\ 1 & 0
\end{matrix}\right),\qquad j_{-}=-\frac{1}{\sqrt{2}}\left(\begin{matrix} 0 & 1 \\ 0 & 0
\end{matrix}\right),\qquad
j_{z}=\frac{1}{2}\left(\begin{matrix}  1 & 0 \\ 0 & -1
\end{matrix}\right),
\end{align}
so that the Killing form is
\begin{align}
\label{FA3}
\mathrm{Tr}\big(j_{a}\hspace{0.1cm}j_{a}\big)=\frac{1}{2}\eta_{a b},\qquad \eta_{ab}=\left(\begin{matrix} 
0 & 1 & 0\\
1 & 0 & 0\\
0 & 0 & 1
\end{matrix}\right),
\end{align}
where the latin indices $a$ and $b$ take the values $+,-,z$. The dreibein $e^{a}{}_{\mu}$ satisfy 
\begin{align}
\label{FA4}
g_{\mu\nu}(x)=e^{a}{}_{\mu}(x)e^{b}{}_{\nu}(x)\eta_{ab},
\end{align}
or, defining the one-forms $e^a=e^a{}_{\mu}\D x^{\mu}$, 
\begin{align}
\label{FA5}
\D s^2=g_{\mu\nu}\D x^{\mu}\D x^{\nu}=\eta_{ab}e^ae^b=(e^z)^2+2e^+e^-.
\end{align}
The Hodge dual of the spin connection $\omega^{ab}=\omega^{ab}{}_{\mu}\D x^{\mu}$ is defined as
\begin{align}
\label{FA6}
\omega^a=-\frac{1}{2}\epsilon^{abc}\omega_{bc},\qquad \epsilon_{z+-}=-\epsilon^{z+-}=1,
\end{align} 
whereas the chiral and anti-chiral connections as
\begin{align}
\label{FA7}
A=\bigg(\omega^a+\frac{e^a}{\ell}\bigg)j_a,\qquad\bar{A}=\bigg(\omega^a-\frac{e^a}{\ell}\bigg)j_a.
\end{align}
The one-forms $e^a$ are chosen to be
\begin{align}
\label{FA8}
e^{\pm}=-\frac{1}{\sqrt{2}}\bigg[\frac{e^{\varphi}}{\rho}\D x^{\pm}-\rho e^{-\varphi} \big(g^{(2)}_{\mp\mp}\D x^{\mp}+g^{(2)}_{+-}\D x^{\pm}\big)\bigg],\qquad e^{z}=-\frac{\ell}{\rho} \D\rho,
\end{align}
and the dual of the spin connection 
\begin{align}
\label{FA9}
\omega^{\pm}=-\frac{1}{\sqrt{2}\ell}\bigg[\frac{e^{\varphi}}{\rho}\D x^{\pm}+\rho e^{-\varphi}\big(g^{(2)}_{\mp\mp}\D x^{\mp}+g^{(2)}_{+-}\D x^{\pm}\big)\bigg],\qquad \omega^{z}=\partial_-\varphi\D x^--\partial_+\varphi\D x^+.
\end{align}
It follows that the left and right connections are given by $A=A_{\mu}\D x^{\mu}$ and $\bar{A}=\bar{A}_{\mu}\D x^{\mu}$
\begin{align}
\label{FA10}
&A_+=\left(\begin{matrix}-\displaystyle{\frac{1}{2}}\partial_+\varphi & \displaystyle{\frac{e^{-\varphi}\rho}{\ell}}g^{(2)}_{++}\\ \\ \displaystyle{\frac{e^{\varphi}}{\ell\rho}} & \displaystyle{\frac{1}{2}}\partial_+\varphi
\end{matrix}\right),\quad A_-=\left(\begin{matrix}\displaystyle{\frac{1}{2}}\partial_-\varphi & \displaystyle{\frac{e^{-\varphi}\rho}{\ell}}g^{(2)}_{+-}\\ \\ 0 & -\displaystyle{\frac{1}{2}}\partial_-\varphi
\end{matrix}\right),\quad A_{\rho}=\left(\begin{matrix}-\displaystyle{\frac{1}{2\rho}}&\hspace{0.5cm} 0\\ \\ 0 &\hspace{0.5cm} \displaystyle{\frac{1}{2\rho}}
\end{matrix}\right),\\\nonumber \\ &\bar{A}_+=\left(\begin{matrix}\displaystyle{-\frac{1}{2}}\partial_+\varphi & 0\\ \\ \displaystyle{\frac{e^{-\varphi}\rho}{\ell}}g^{(2)}_{+-} & \displaystyle{\frac{1}{2}}\partial_+\varphi
\end{matrix}\right),\quad \bar{A}_-=\left(\begin{matrix}\displaystyle{\frac{1}{2}}\partial_-\varphi & \displaystyle{\frac{e^{\varphi}}{\ell\rho}} \\ \\ \displaystyle{\frac{e^{-\varphi}\rho}{\ell}}g^{(2)}_{--} & -\displaystyle{\frac{1}{2}}\partial_-\varphi
\end{matrix}\right),\quad \bar{A}_{\rho}=\left(\begin{matrix}\displaystyle{\frac{1}{2\rho}}&\hspace{0.5cm} 0\\ \\ 0 &\hspace{0.5cm} -\displaystyle{\frac{1}{2\rho}}
\end{matrix}\right).
\end{align}
Note that with BH boundary conditions, $\varphi=0$, $A_{+}$ is chiral, $A_-=0$ and $\bar{A}_-$ is anti-chiral, $\bar{A}_{+}=0$.

\subsection{Variational Problem, Weyl Anomaly and WZW Reduction}

Let us now discuss the action principle and the variatonal problem associated with \eqref{FA1}. We find it convenient to discuss it in terms of  coordinates $(\rho,t,\phi)$. The action contains a pure boundary term that does not change the dynamics and that we ignore. Indeed, we define our starting action as (the dot indicates a $t$ derivative while prime a $\phi$ derivative)
\begin{align}
\label{FA11}
\tilde{S}_{CS}[A]=-\kappa\int_{M}\D^3x\mathrm{Tr}\big(A_{\rho}\dot{A}_{\phi}-A_{\phi}\dot{A}_{\rho}+2A_tF_{\phi\rho}\big).
\end{align}
Taking a variation of \eqref{FA11} yields
\begin{align}
\nonumber\delta \tilde{S}_{CS}[A]=-\kappa\int_{M}\D^3x\text{Tr}\big(2\delta A_rF_{t\phi}-2\delta A_{\phi}F_{tr}+2\delta A_tF_{\phi r}\big)+2\kappa\int_{\partial M}\D^2x\text{Tr}\big(A_t\delta A_{\phi}\big)=2\kappa\int_{\partial M}\D^2x\text{Tr}\big(A_t\delta A_{\phi}\big),
\end{align}
where in the last step we have imposed the equations of motion, $F=dA=0$. In total, considering also the contribution of the anti-chiral sector
\begin{align}
\label{FA13}
\delta \tilde{S}_{CS}[A]-\delta \tilde{S}_{CS}[\bar{A}]= 2\kappa \int_{\partial M}\D^2x\mathrm{Tr}\big(A_t\delta A_{\phi}-\bar{A}_t\delta\bar{A}_{\phi}\big).
\end{align}

With BH boundary conditions, in order to have a well-defined variational problem, it is sufficient to add to the action the Coussaert--Henneaux--Van Driel boundary term \cite{Coussaert:1995zp},
\begin{align}
\label{FA14}
\tilde{S}[A,\bar{A}]=\tilde{S}_{CS}[A]-\tilde{S}_{CS}[\bar{A}]-\frac{\kappa}{\ell}\int_{\partial M}\D^2x \mathrm{Tr}\big(A^2_{\phi}+\bar{A}^2_{\phi}\big),
\end{align}
whose variation cancels exactly the right-hand side of \eqref{FA13}, since on shell $A_t=\frac{1}{\ell}A_{\phi}$ and $\bar{A}_t=-\frac{1}{\ell}\bar{A}_{\phi}$. However, with our choice of boundary conditions the variation of the action is
\begin{align}
\label{FA15}
\delta \tilde{S}[A,\bar{A}]=-\kappa\ell \delta\int_{\partial M}\D^2 x (\partial_t\varphi)^2+\frac{2\kappa}{\ell}\int_{\partial M}\D^2 x(\ell^2\partial^2_t-\partial^2_{\phi})\varphi\delta\varphi.
\end{align}
The first term is already integrated while the last term is not integrable, due to the Weyl anomaly.
With the decomposition 
\begin{align}
\label{FA16}
A_{\mu}=A_{\mu}^aj_a,\Longrightarrow A_{\phi}^{z}=\bar{A}_{\phi}^{z}=-\ell\partial_{t}\varphi,
\end{align}
we can write the action as
\begin{align}
\label{FA17}
S[A,\bar{A}]=\tilde{S}_{CS}[A]-\tilde{S}_{CS}[\bar{A}]-\frac{\kappa}{\ell}\int_{\partial M}\D^2x\mathrm{Tr}\big(A^2_{\phi}+\bar{A}^2_{\phi}\big)+\frac{\kappa}{\ell}\int_{\partial M}\D^2x A_{\phi}^{z}\bar{A}^z_{\phi}.
\end{align}
The variational problem for this action is ill-defined, for the theory is Weyl anomalous. In other words, it is not possible to add more boundary terms to the action to achieve $\delta S=0$.  

Let us now perform the reduction to a WZW model \cite{Wess:1971yu, Witten:1983tw, Witten:1983ar, Elitzur:1989nr}. Solving the constraints, the spatial components of the connection are given by
\begin{align}
\label{FA20}
A_i=G^{-1}\partial_i G,\hspace{1cm}\bar{A}_i=\bar{G}^{-1}\partial_i\bar{G},
\end{align}
for some elements $G\in\mathrm{SL}(2,\mathbb{R})_L$ and  $\bar{G}\in\mathrm{SL}(2,\mathbb{R})_R$. The constraints $F_{\rho \phi}=0$ and $\bar{F}_{\rho \phi}=0$ imply that $G$ and $\bar{G}$ have the form
\begin{align}
\label{FA21}
G=g(t,\phi)h(\rho,t),\qquad\bar{G}=\bar{g}(t,\phi)\bar{h}(\rho,t),
\end{align}
as can be easily verified. Furthermore we assume that $\partial_t h(\rho,t)|_{\partial M}=\partial_t \bar{h}(\rho,t)|_{\partial M}=0$. Plugging this into $S[A,\bar{A}]$, we have, after some algebra,  
\begin{align}
\label{FA22}
\nonumber
\tilde{S}[A,\bar{A}]=&\frac{\kappa}{\ell}\int_{\partial M}\D^2x\mathrm{Tr}\big[g^{-1}\partial_{\phi} g(\ell^{-1}g^{-1}\partial_t g-g^{-1}\partial_{\phi}g)\big]+\frac{\kappa}{3}\int_{M}\mathrm{Tr}\big(G^{-1}dG\wedge G^{-1}dG\wedge G^{-1}dG\big)\\\nonumber-&\frac{\kappa}{\ell}\int_{\partial M}\D^2x\mathrm{Tr}\big[\bar{g}^{-1}\partial_{\phi} \bar{g}(\ell^{-1}\bar{g}^{-1}\partial_t \bar{g}+\bar{g}^{-1}\partial_{\phi}\bar{g})\big]-\frac{\kappa}{3}\int_{M}\mathrm{Tr}\big(\bar{G}^{-1}d\bar{G}\wedge \bar{G}^{-1}d\bar{G}\wedge \bar{G}^{-1}d\bar{G}\big)\\+&\frac{\kappa}{\ell}\int_{\partial M}\D^2xA_{\phi}^z\bar{A}^z_{\phi}.
\end{align}
This is the WZW reduced action.

\subsection{Gauss Decomposition}

Let us focus on the chiral part of the action \eqref{FA22} and consider the following decomposition of $g$
\begin{align}
\label{FA23}
g=\left(\begin{matrix}
1 & 0\\ \sigma & 1
\end{matrix}\right)\left(\begin{matrix}
e^{-\chi/2} & 0\\ 0 & e^{\chi/2}
\end{matrix}\right)\left(\begin{matrix}
1 & \tau\\ 0 & 1
\end{matrix}\right)=\left(\begin{matrix}
e^{-\chi/2} & \tau e^{-\chi/2}\\ \sigma e^{-\chi/2} & \sigma \tau e^{-\chi/2}+e^{\chi/2}
\end{matrix}\right),
\end{align}
from which it follows that
\begin{align}
\label{FA24}
g^{-1}\partial_{\mu}g=\left(\begin{matrix}&-e^{-\chi}\tau\partial_{\mu}\sigma-\frac{1}{2}\partial_{\mu}\chi & -e^{-\chi}\tau^2\partial_{\mu}\sigma+\partial_{\mu}\tau-\tau\partial_{\mu}\chi \\ &e^{-\chi}\partial_{\mu}\sigma & e^{-\chi}\tau\partial_{\mu}\sigma+\frac{1}{2}\partial_{\mu}\chi
\end{matrix}\right).
\end{align}
In terms of the Gauss fields $(\sigma,\chi,\tau)$, the boundary term is
\begin{align}
\label{FA25}
\frac{\kappa}{\ell}\int_{\partial M}\D^2x\mathrm{Tr}\big[g^{-1}\partial_{\phi} g(\ell^{-1}g^{-1}\partial_t g-g^{-1}\partial_{\phi}g)\big]=\frac{\kappa}{\ell}\int_{\partial M}\D^2x\big[\frac{1}{2}\chi'(\ell\dot{\chi}-\chi')+\ell e^{-\chi}(\tau'\dot{\sigma}+\dot{\tau}\sigma')-2e^{-\chi}\tau'\sigma'\big].
\end{align}
For the bulk term first note that, decomposing $G$ as
\begin{align}
\label{FA26}
G=\left(\begin{matrix}
1 & 0\\ \Sigma & 1
\end{matrix}\right)\left(\begin{matrix}
e^{-X/2} & 0\\ 0 & e^{X/2}
\end{matrix}\right)\left(\begin{matrix}
1 & T\\ 0 & 1
\end{matrix}\right),
\end{align}
we gather
\begin{align}
\label{FA27}
\mathrm{Tr}\big[G^{-1}dG\wedge G^{-1}dG\wedge G^{-1}d G\big]=-3\epsilon^{\mu\nu\lambda}\partial_{\mu}\big(e^{-X}\partial_{\nu}\Sigma\partial_{\lambda}T\big)\D\rho\wedge\D t\wedge\D\phi.
\end{align}
Hence, applying Stokes theorem,
\begin{align}
\label{FA28}
\frac{\kappa}{3}\int_{M}\mathrm{Tr}[G^{-1}dG\wedge G^{-1}dG\wedge G^{-1}dG]=-\kappa\int_{\partial M}\D^2x e^{-X}(\dot{\Sigma}T'-\Sigma'\dot{T})\big |_{\partial M}.
\end{align}
Furthermore, for the matrix $h(\rho,t)$ we have
\begin{align}
\label{FA29}
A_{\rho}=h^{-1}\partial_{\rho} h=\left(\begin{matrix}-\frac{1}{2\rho} & 0\\ \\ 0 &\hspace{0.5cm} \frac{1}{2\rho}\end{matrix}\right)\Longrightarrow h=\left(\begin{matrix}
\sqrt{\frac{\ell}{\rho}} & 0\\ 0 &\sqrt{\frac{\rho}{\ell}}
\end{matrix}\right).
\end{align}
Since $G=g(t,\phi) h(\rho,t)$ we have the equality 
\begin{align}
\label{FA30}
\left(\begin{matrix}
e^{-X/2} & T e^{-X/2}\\ \Sigma e^{-X/2} & \Sigma T e^{-X/2}+e^{X/2}
\end{matrix}\right)=\left(\begin{matrix}&e^{-\chi/2}\sqrt{\frac{\ell}{\rho}}, &e^{-\chi/2}\sqrt{\frac{\rho}{\ell}}\tau \\& e^{-\chi/2}\sqrt{\frac{\ell}{\rho}}\sigma & e^{-\chi/2}\sqrt{\frac{\rho}{\ell}}(e^{\chi}+\sigma\tau)
\end{matrix}\right),
\end{align}
which gives
\begin{align}
\label{FA31}
e^{-X}=\frac{\ell}{\rho}e^{-\chi},\hspace{1cm}T=\frac{\rho}{\ell}\tau,\hspace{1cm}\Sigma=\sigma.
\end{align}
Hence the term integral in \eqref{FA28} is $-\kappa\int_{\partial M}\D^2x e^{-\chi}(\dot{\sigma}\tau'-\sigma'\dot{ \tau})$ in terms of Gauss fields. The full chiral part of the action thus reads
\begin{align}
\label{FA31}
\frac{\kappa}{\ell}\int_{\partial M}\D^2x\big[\frac{1}{2}\chi'(\ell\dot{\chi}-\chi')+2e^{-\chi}\sigma'(\ell\dot{\tau}-\tau ')\big].
\end{align}
Assuming for $\bar{g}$ and $\bar{G}$ the decompositions
\begin{align}
\label{FA32}
& \bar{g}=\left(\begin{matrix}
1 & \bar{\sigma}\\ 0 & 1
\end{matrix}\right)\left(\begin{matrix}
e^{\bar{\chi}/2} & 0\\ 0 & e^{-\bar{\chi}/2}
\end{matrix}\right)\left(\begin{matrix}
1 & 0\\ \bar{\tau} & 1
\end{matrix}\right)=\left(\begin{matrix}
 \bar{\sigma}\bar{\tau} e^{-\bar{\chi}/2}+e^{\bar{\chi}/2} & \bar{\sigma} e^{-\bar{\chi}/2}\\ \bar{\tau} e^{-\bar{\chi}/2} & e^{-\bar{\chi}/2}
\end{matrix}\right),\\
&\bar{G}=\left(\begin{matrix}
1 & \bar{\Sigma}\\ 0 & 1
\end{matrix}\right)\left(\begin{matrix}
e^{\bar{X}/2} & 0\\ 0 & e^{-\bar{X}/2}
\end{matrix}\right)\left(\begin{matrix}
1 & 0\\ \bar{T} & 1
\end{matrix}\right)=\left(\begin{matrix}
 \bar{\Sigma}\bar{T} e^{-\bar{X}/2}+e^{\bar{X}/2} & \bar{\Sigma} e^{-\bar{X}/2}\\ \bar{T} e^{-\bar{X}/2} & e^{-\bar{X}/2}
\end{matrix}\right),
\end{align}
and noting that $\bar{h}(\rho,t)$ is given by
\begin{align}
\label{FA33}
\bar{h}=\left(\begin{matrix}
\sqrt{\frac{\rho}{\ell}} & 0\\ 0 & \sqrt{\frac{\ell}{\rho}}
\end{matrix}\right).
\end{align}
The procedure used for the chiral part can be repeated for the anti-chiral part of the action so that
\begin{align}
\label{FA34}
\frac{\kappa}{\ell}\int_{\partial M}\D^2x\big[-\frac{1}{2}\bar{\chi}'(\ell\dot{\bar{\chi}}+\bar{\chi}')-2e^{-\bar{\chi}}\bar{\sigma}'(\ell\dot{\bar{\tau}}+\bar{\tau}')\big].
\end{align}
The total action in terms of the Gauss fields is then
\begin{align}
\label{FA35}
S[A,\bar{A}]=\frac{\kappa}{\ell}\int_{\partial M}\D^2x\big[\frac{1}{2}\chi'(\ell\dot{\chi}-\chi')+2e^{-\chi}\sigma'(\ell\dot{\tau}-\tau ')-\frac{1}{2}\bar{\chi}'(\ell\dot{\bar{\chi}}+\bar{\chi}')-2e^{-\bar{\chi}}\bar{\sigma}'(\ell\dot{\bar{\tau}}+\bar{\tau}')+A_{\phi}^z\bar{A}_{\phi}^z\big].
\end{align}
Note that it is possible to express $A_{\phi}^{z}$ and $\bar{A}^{z}_{\phi}$ in terms of the Gauss fields as
\begin{align}
\label{FA36}
A_{\phi}^{z}=-2e^{-\chi}\tau\sigma'-\chi'=\bar{A}_{\phi}^{z}=2e^{-\bar{\chi}}\bar{\tau}\bar{\sigma}'+\bar{\chi}'.
\end{align}
Defining
\begin{align}
\label{FA36.5} C=\frac{A^{z}_{\phi}}{2}=\frac{\bar{A}^{z}_{\phi}}{2}=-\frac{\ell}{2}\partial_t\varphi.
\end{align}
 we can rewrite the mixing term as a quadratic term $4C^2$. 
We perform now the Hamiltonian analysis of the action \eqref{FA35}.
The canonical momenta $\pi_{\phi_i}=\frac{\partial \mathcal{L}}{\partial \dot{\phi}^i}$ are 
\begin{align}
\label{FA37}
&\pi_{\chi}=\frac{\kappa}{2}\chi',\hspace{1cm}\pi_{\tau}=2\kappa e^{-\chi}\sigma',\hspace{1cm}\pi_{\sigma}=0,\\
\label{FA37.5}
&\pi_{\bar{\chi}}=-\frac{\kappa}{2}\bar{\chi}',\hspace{0.7cm}\pi_{\bar{\tau}}=-2\kappa e^{-\bar{\chi}}\bar{\sigma}',\hspace{0.7cm}\pi_{\bar{\sigma}}=0,
\end{align}
together with $\pi_{C}=0$. The Hamiltonian density is
\begin{align}
\label{FA38}
\mathcal{H}=\dot{\phi}^{i}\pi_{\phi_i}-\mathcal{L}=\frac{\kappa}{\ell}\bigg(\frac{1}{2}\chi'^2+\frac{1}{2}\bar{\chi}'^2+
2e^{-\chi}\sigma'\tau'+2e^{-\bar{\chi}}\bar{\sigma}'\bar{\tau}'-4C^2\bigg).
\end{align}
Now we implement our boundary conditions, using the equalities $g^{-1}\partial_{\phi}g=hA_{\phi}h^{-1}\big|_{\partial M}$ and $\bar{g}^{-1}\partial_{\phi}\bar{g}=\bar{h}\bar{A}_{\phi}\bar{h}^{-1}\big|_{\partial M}$. We obtain the following relations
\begin{align}
\label{FA38.5}
&\bullet C=-e^{-\chi}\tau\sigma'-\frac{1}{2}\chi'=e^{-\bar{\chi}}\bar{\tau}\bar{\sigma}'+\frac{1}{2}\bar{\chi}',\\ 
\label{FA38.6}
&\bullet e^{-\chi}\sigma'=\displaystyle{\frac{e^{\varphi}}{\ell^2}}=-e^{-\bar{\chi}}\bar{\sigma}',\\
\label{FA38.7}
&\bullet-e^{-\chi}\tau^2\sigma'+\tau'-\tau\chi'=e^{-\varphi}(g^{(2)}_{++}-g^{(2)}_{+-})\\
\label{FA38.8}
&\bullet -e^{-\bar{\chi}}\bar{\tau}^2\bar{\sigma}'-\bar{\tau}'+\bar{\tau}\bar{\chi}'=-e^{-\varphi}(g^{(2)}_{--}-g^{(2)}_{+-}).
\end{align}
Plugging them in the hamiltonian of \eqref{FA38} we have
\begin{align}
\label{FA39}
\mathcal{H}=\frac{\kappa}{\ell}\bigg(\frac{1}{2}\chi'^2+\frac{1}{2}\bar{\chi}'^2-\chi''-\bar{\chi}''+\varphi'(\chi'+\bar{\chi}')-4C^2\bigg),
\end{align}
where we note that $C$ cannot be further expressed in terms of the other independent fields. Note also that the Hamiltonian can be simply expressed as
\begin{align}
\label{FA40}
\mathcal{H}=\frac{2\kappa}{\ell^3}\big(g^{(2)}_{++}+g^{(2)}_{--}-2g^{(2)}_{+-}\big)=\ell T_{tt},
\end{align}
as it is reasonable, using \eqref{FA38.5}-\eqref{FA38.8}. Let us turn to the equations of motion. The Hamiltonian action is
\begin{align}
\label{FA41}
S_{H}=\int_{\partial M}\D^2x \big(\pi_{\chi}\dot{\chi}+\pi_{\bar{\chi}}\dot{\bar{\chi}}+\pi_{\tau}\dot{\tau}+\pi_{\bar{\tau}}\dot{\bar{\tau}}-\mathcal{H}\big),
\end{align}
and, using equations \eqref{FA37} and \eqref{FA37.5}, together with the relations \eqref{FA38.5}-\eqref{FA38.8}, we get
\begin{align}
\label{FA42}
S_{H}&=\frac{\kappa}{\ell}\int_{\partial M}\D^2 x\big[\big(\frac{1}{2}\chi'+\varphi'\big)(\ell\dot{\chi}-\chi')-\big(\frac{1}{2}\bar{\chi}'+\varphi'\big)(\ell\dot{\bar{\chi}}+\bar{\chi}')+4\ell C\dot{\varphi}+4C^2\big].
\end{align}
It follows from \eqref{FA42} that $C$ is proportional to the canonical momentum conjugate to $\varphi$. The Poisson bracket 
\begin{align}
\label{FA43}
\{\varphi(t,\phi),C(t,\phi')\}=\frac{1}{4\kappa}\delta(\phi-\phi'),\Longrightarrow \{\varphi(t,\phi),\partial_t\varphi(t,\phi')\}=-\frac{8\pi G}{\ell^2}\delta(\phi-\phi')
\end{align}
where we have used \eqref{FA36.5} to express $C$ in terms of $\partial_t\varphi$. Using again \eqref{FA36.5} in \eqref{FA43}, we get
\begin{align}
\label{FA42.3}
S_{H}=\frac{\kappa}{\ell}\int_{\partial M}\D^2 x\big[\big(\frac{1}{2}\chi'+\varphi'\big)(\ell\dot{\chi}-\chi')-\big(\frac{1}{2}\bar{\chi}'+\varphi'\big)(\ell\dot{\bar{\chi}}+\bar{\chi}')-\ell^2\dot{\varphi}^2\big].
\end{align}
The action \eqref{FA42.3} mixes $\chi$ and $\bar{\chi}$ with $\varphi$, but it is straightforward to show that, introducing new fields
\begin{align}
\label{FA45}
\psi=\chi+\varphi,\qquad \bar{\psi}=\bar{\chi}+\varphi,
\end{align}
it admits a simple rewriting
\begin{align}
\label{FA46}
S_{H}=\frac{\kappa}{\ell}\int_{\partial M}\D^2x\big[\frac{1}{2}\psi'(\ell\partial_t-\partial_{\phi})\psi-\frac{1}{2}\bar{\psi}'(\ell\partial_t+\partial_{\phi})\bar{\psi}-\ell^2(\partial_t\varphi)^2+(\partial_{\phi}\varphi)^2\big].
\end{align}
From \eqref{FA46} it is clear that the dynamics of $\psi$ and $\bar{\psi}$ is independent of $\varphi$, which is the desired result. The action \eqref{FA46} can be shown to be equivalent to a Liouville theory \cite{Forgacs:1989ac,Carlip:2001kk,Poojary:2017xgn,Donnay:2016iyk} coupled to an external two-dimensional metric in conformal gauge. Eventually, we stress again that the Chern--Simons construction carried out so far shows that the $\varphi$ reduced boundary action (i.e. the last terms in \eqref{FA46}) is completely disentangled from the rest. 

\providecommand{\href}[2]{#2}\begingroup\raggedright\endgroup
\end{document}